\documentclass[aps,nofootinbib,superscriptaddress, showpacs,preprintnumbers,  nofootinbibt,twocolumn]{revtex4-2}
\usepackage{amsmath,amsfonts,amssymb,graphicx}
\usepackage{xcolor}
\usepackage{multirow}
\usepackage{float}
\usepackage{rotating}
\usepackage[colorlinks,citecolor={violet}]{hyperref}
\def\be{\begin{equation}}
\def\ee{\end{equation}}
\def\bea{\begin{eqnarray}}
\def\eea{\end{eqnarray}}
\begin{document}
\title{Dark energy and accelerating cosmological evolution in a Universe with a Weylian boundary}
\author{Tiberiu Harko}
\email{tiberiu.harko@aira.astro.ro}
\affiliation{Faculty of Physics, Babe\c s-Bolyai University, 1 Kog\u alniceanu Street,	400084 Cluj-Napoca, Romania,}
\affiliation{Astronomical Observatory, 19 Cire\c silor Street, 400487, Cluj-Napoca, Romania,}
\author{Shahab Shahidi}
\email{s.shahidi@du.ac.ir}
\affiliation{School of Physics, Damghan University, Damghan, 36716-45667, Iran}
\date{\today }

\begin{abstract}
We investigate the influence of boundary terms in gravitational field theories, by considering that in the Einstein-Hilbert action the boundary can be described by a non-metric Weyl-type geometry. The gravitational action and the the field equations, are thus generalized to include new geometrical terms, coming from the non-metric nature of the boundary, and depending on the Weyl vector, and its covariant derivatives. The field equations obtained within this framework  generalize the standard Einstein equations by including in their mathematical structure the Weyl vector, and its covariant derivatives. As an applications of the general formalism we investigate the cosmological evolution in a flat FLRW geometry. We obtain the generalized Friedmann equations, which contain extra terms depending on the Weyl vector and its derivatives, arising due to the presence of the Weylian boundary, and which  describe an effective, time dependent dark energy. By imposing to the dark energy an equation of state parameter of the Barboza-Alcaniz type, the Friedmann equations can be solved numerically. We compare the predictions of the Weylian boundary gravitational theory with late-time observational data and the predictions of the $\Lambda$CDM paradigm. Our results show that the Weylian boundary cosmological models give a good description of the observational data, and they can reproduce almost exactly the predictions of the $\Lambda$CDM paradigm. Hence, the extension of gravitational theories through the addition of Weylian boundary terms, in which dark energy has a purely geometric origin, emerges as a viable alternative to standard general relativity.   
\end{abstract}
\maketitle

\tableofcontents

\section{Introduction}

A long standing problem in general relativity is the boundary value problem of the Einstein-Hilbert variational principle. The problem is related to the necessity to guaranty that the initial conditions for the evolution of gravitational systems are well defined. To solve this problem  in \cite{GH, Y}  an additional term was added to the Einstein-Hilbert action, relating the boundary and the extrinsic curvature of the space-time. On the other hand, it is a standard result that a total derivative term can be obtained from the Ricci scalar, leading to the so-called  Gamma
squared action \cite{Bo1,Bo2,Bo3}. This action also gives rise to the Einstein field equations when variations with respect
to the metric are considered. However, the presence of the boundary term, when  coupled with a scalar field and a cosmological fluid, can have a significant impact on the cosmological evolution \cite{Bo3}. 

An alternative proposal for the Gibbons-Hawking-York term was considered in \cite{BR1, BR2}, and it is based on the assumption that the boundary term must be considered as a true physical source having a geometric nature. The influence of the boundary terms was considered in a Weyl type geometry, considered to exist in the preinflationary Universe. In this approach the variation of the Ricci curvature tensor is introduced according to  the definition $\delta R_{\alpha \beta}\equiv \nabla_{\mu} W^{\mu}=\phi(x^{\alpha})$, where $W$ is a tetra-vector of geometric kind, and $\phi$ is a scalar field. The tetra-vector field is gauge invariant under the transformation $\delta \tilde W_\alpha = \delta W_\alpha - \Lambda g_{\alpha \beta }$ \cite{BR1,BR2}. The definition of the generalized Einstein tensor $\tilde G_{\alpha \beta}$ includes  the boundary parameter $\Lambda$, and thus $\tilde G_{\alpha \beta}=G_{\alpha \beta }-\Lambda g_{\alpha \beta}$. Hence, the field equations obtained within this approach to the presence of boundary terms are $G_{\alpha \beta }+\Lambda g_{\alpha \beta}=\kappa T_{\alpha \beta}$. For the case of an FLRW metric, the role of the boundary term was analyzed  in \cite{B3}, by assuming that the matter source can be described by a barotropic fluid. This assumption allowed the determination of the expression of the cosmological boundary flux parameter $\Lambda $.

In the framework of the $f(Q)$ gravity \cite{fq1,fq2,fq3,fq4}, where $Q$ is the non-metricity, the role of the boundary terms  was considered in  \cite{DLS}. The inclusion of a boundary led to the formulation of the $f(Q,C)$ theory, with the boundary term $C$ given by the difference between the standard Levi-Civita Ricci scalar  $\mathring{R}$ and $Q$, $C=\mathring{R}-Q$. The cosmological implications of the theory were also investigated, and it was shown that effective dark-energy sector of geometrical origin can be obtained. An effective interaction between matter and dark energy is also generated due to the presence of the boundary terms. Cosmological inflation in the $f(Q,C)$ modified gravity theory was investigated in \cite{fQC1}. For other investigations of the $f(Q,C)$ theoretical approach see \cite{fQC2,fQC3,fQC4,fQC5,fQC6,fQC7}.

 Another modified gravity theory which considers the role of the boundary terms is called $f(Q,B)$ gravity, and it was considered in \cite{CFF}.  The field equations were derived by using a variational principle, and compared with the field equations of $f(Q)$ gravity, obtained in the limit $B\rightarrow 0$. An interesting property of these theories is that the  $f(Q,B)=f(Q-B)$ models are dynamically equivalent to $f(R)$ gravity. This situation is similar to the teleparallel $f(\tilde{B}-T)$ gravity, where $\tilde{B}\neq B$. For further investigations on the role of boundary in modified gravity theories see \cite{Kad,Pal, Loh}. 
 
 In \cite{Harko} the influence of boundary terms in the warm inflationary scenario were investigated, by considering that in the Einstein–Hilbert action the boundary can be described in terms of a Weyl-type geometry. The gravitational action, as well as the field equations, have been generalized to include new geometrical terms, coming from the non-metric nature of the boundary, and depending on the Weyl vector, and its covariant derivatives. The effects of the Weylian boundary terms were investigated by considering the warm inflationary scenario of the early evolution of the Universe, in the presence of a scalar field. It turns out that the Weyl vector, describing the boundary effects on the cosmological evolution, plays a significant role during the process of radiation creation. 

It is the goal of the present investigation to generalize the approach initiated in \cite{Harko} to a cosmological scenario in which dark energy, which triggered the accelerating expansion of the Universe, is generated by the boundary terms in the Einstein-Hilbert action. As for the nature of the boundary, we assume that it is described by a Weyl type non-metric geometry, in which the covariant derivative of the metric tensor does not vanish, $\nabla_{\mu}g_{\alpha \beta} \neq 0$. 

In Weyl geometry, non-metricity emerges naturally due to the presence of the Weyl gauge vector field, $\omega_{\mu}$.  From a theoretical point of view we begin our analysis by extending  the Einstein-Hilbert action principle so that the boundary terms are described  in the Weyl geometric framework. By varying the action with respect to the metric tensor we obtain the gravitational field equations containing the contributions of the Weylian boundary terms. The generalized Friedmann equations are obtained for a flat FLRW geometry, and they contain extra terms, originating from the boundary,  which depend on the Weyl vector, and its time derivative, and which we interpret as describing an effective, time varying dark energy. 

By introducing a set of dimensionless variables, and the redshift representation,  the cosmological implications of the theory are considered for a specific choice of the parameter of the dark energy equation of state. The theoretical predictions of the model are compared with recent observational data, and with the results of the $\Lambda$CDM model. The Weylian boundary gravity theory gives a good description of the observational data, and reproduces almost exactly the predictions of the $\Lambda$CDM model. 

The present paper is organized as follows. The fundamentals of Weyl geometry, the Einstein-Hilbert action principle, and the boundary term are presented in Section~\ref{field}. The action and the derivation of the field equations in the presence of Weylian  boundary terms taking values in the Weyl geometry is presented in Section \ref{derfield}. The implications of the Weylian boundary gravity theory for the understanding of the cosmological properties of the present day Universe are presented in Section~\ref{cosm}. We discuss and conclude our results in Section~\ref{concl}.

\section{Weyl geometry, Einstein-Hilbert action, and boundary terms} \label{field}

In the present Section we review the basics of the Weyl geometry necessary in the sequel, as well as the Hilbert-Einstein variational principle, and the boundary term it generates. We will use these theoretical tools to formulate a geometrodynamical approach that extends standard general relativity by including the effects of the boundary term in the gravitational field equations. The presence of the Weylian boundary terms significantly affect the cosmological dynamics of the Universe. In the present work we use the Landau-Lifshitz \cite{LaLi} conventions for the metric signature, and for the definitions of the geometrical and quantities.

\subsection{Fundamentals of Weyl geometry}

The Weyl geometry \cite{Weyl} is essentially defined by the special role played by conformal transformations, and conformal symmetry. Conformal symmetry is implemented via the requirement  of the invariance of the geometrical (and physical) quantities with respect to the Weyl gauge symmetry transformation
$ \hat g_{\mu\nu}=\Sigma^n(x) \,g_{\mu\nu}$, where $\Sigma(x)>0$, and $n$ is called the Weyl charge. With respect to these transformations $g=\det g_{\mu\nu}$, and the physical fields  $\phi$ and $\psi$,  describing
real bosonic or fermionic particles, transform as \cite{Ghil,Ghil1,Ghil2,Ghil3}
$$ \sqrt{-\hat g} =\Sigma^{2 n} \sqrt{-g},\quad \hat \phi = \Sigma^{-n/2} \phi,\quad\hat\psi=\Sigma^{-3n/4}\,\psi.$$  In the following we consider, without any loss of generality, $n=1$.

The requirement of the conformal invariance can be naturally implemented via the use of the mathematical approach represented by Weyl geometry. Weyl
geometry can be defined in a very general sense as the classes of equivalence $(g_{\alpha \beta },\omega _{\mu }$) of the metric $g_{\mu \nu }$, assumed to be symmetric, and the Weyl vector gauge
field $\omega _{\mu }$ \cite{Del}. The metrics are related via the Weyl gauge symmetry transformation introduced above. These transformations must be extended by the addition of the
transformation rule of the Weyl vector $\omega _{\mu }$, which, in order to maintain gauge invariance, is  given by
\begin{equation}\label{WGS}
\hat{\omega}_{\mu }=\omega _{\mu }-\frac{1}{\alpha }\,\partial _{\mu }\ln
\Sigma .
\end{equation}%
Here the constant $\alpha $ denotes the Weyl gauge coupling parameter. The basic geometrical
property of the Weyl geometry is its non-metric character, which implies that the covariant
derivative of the metric tensor is non-zero. The non-vanishing of the divergence of the metric tensor is related to an important geometric quantity, the non-metricity, determined by
the presence of the Weyl gauge field $\omega _{\mu }$. The non-metricity is defined with the help of the relation \cite{Ghil,Ghil1,Ghil2,Ghil3, Del}
\begin{equation}\label{nm}
\tilde{\nabla}_{\mu }g_{\alpha \beta }=-\alpha \omega _{\mu }g_{\alpha \beta
}.  
\end{equation}%

Equation (\ref{nm}) gives the connection $\tilde{\Gamma}$ of the Weyl geometry as  \cite{Ghil,Ghil1, Ghil2,Ghil3, Del}
\begin{align}
\tilde{\Gamma}_{\mu \nu }^{\lambda } =&\Gamma
_{\mu \nu }^{\lambda }+\Psi _{\mu \nu }^{\lambda }\nonumber\\=&\Gamma _{\mu \nu }^{\lambda }+\frac{1%
}{2}\alpha \big(\delta _{\mu }^{\lambda }\,\,\omega _{\nu }+\delta _{\nu
}^{\lambda }\,\,\omega _{\mu }-g_{\mu \nu }\,\omega ^{\lambda }\big),  \notag  \label{tGamma}
\end{align}%
where $\Gamma _{\mu \nu }^{\lambda }$ denotes the Levi-Civita connection, obtained
through its standard definition
\be
\Gamma _{\mu \nu }^{\alpha }(g)=\frac{1}{2}g^{\alpha
\lambda }(\partial _{\mu }g_{\lambda \nu }+\partial _{\nu }g_{\lambda \mu
}-\partial _{\lambda }g_{\mu \nu }),
\ee
while $\Psi _{\mu \nu }^{\lambda}$, giving the contribution to the connection coming from the Weyl vector, is given by
\be
\Psi _{\mu \nu }^{\lambda
}=\frac{\alpha}{2}\Big(\delta _{\mu }^{\lambda }\,\,\omega _{\nu }+\delta _{\nu
}^{\lambda }\,\,\omega _{\mu }-g_{\mu \nu }\,\omega ^{\lambda }\Big).
\ee

 By contracting the connection coefficients we find $\tilde{\Gamma}%
_{\mu \nu }^{\nu }=\tilde{\Gamma}_{\mu }$, and $\Gamma _{\mu \nu }^{\nu
}=\Gamma _{\mu }$, respectively. With the use of the contractions of the connection, the Weyl vector is obtained as
\begin{equation}
\omega _{\mu }=\frac{1}{2\alpha }\left( \tilde{\Gamma}_{\mu }-\Gamma _{\mu
}\right) .
\end{equation}

Hence, the Weyl gauge vector $\omega _{\mu }$ can be interpreted as giving the deviation of the trace of the Weyl
connection from the Levi-Civita connection \cite{Ghil}. Since $%
\omega _{\mu }$ is a component of the Weyl connection $\tilde{\Gamma}$,  it follows that it is
a purely geometric quantity. 

The covariant derivative of a typical vector
is obtained in Weyl geometry according to the prescription
\begin{align}\label{cov1}
	\tilde{\nabla}_{\mu }\omega _{\nu } &=\frac{\partial \omega _{\nu }}{%
		\partial x^{\nu }}-\tilde{\Gamma}_{\mu \nu }^{\lambda }\omega _{\lambda }=%
	\frac{\partial \omega _{\nu }}{\partial x^{\nu }}-\Gamma _{\mu \nu
	}^{\lambda }\omega _{\lambda }-\Psi _{\mu \nu }^{\lambda }\omega _{\lambda }\nonumber\\
	&=\nabla _{\mu }\omega _{\nu }-\Psi _{\mu \nu }^{\lambda }\omega _{\lambda
	}%\nonumber\\
	=\nabla _{\mu }\omega _{\nu }-\alpha \omega _{\mu }\omega _{\nu }+\frac{1}{2%
	}\alpha g_{\mu \nu }\omega _{\lambda }\omega ^{\lambda }.\nonumber\\
\end{align}
Also we have
\begin{align}\label{cov2}
\tilde{\nabla}_{\mu}\omega ^{\lambda }&=\nabla _{\mu}\omega
^{\lambda }+\frac12\alpha \delta^\lambda_\mu \omega _{\alpha }\omega ^{\alpha }.\\
\tilde{\nabla}_{\lambda }\omega ^{\lambda }&=\nabla _{\lambda }\omega
^{\lambda }+2\alpha \omega _{\lambda }\omega ^{\lambda }.
\end{align}

The Weyl connection $\tilde{\Gamma}$ is invariant with respect to the
combined set of gauge symmetry transformations  of the metric $g_{\mu\nu}$ and of the gauge field $\omega _\mu$. In a special case where $\omega _{\mu
}$ can be represented in a pure gauge form, as the gradient of a scalar quantity,  then the obtained 
geometry is called the Weyl integrable geometry \cite{WIG}. The Weyl integrable geometry is a metric geometry, but it is still more general than the Riemannian one.

We can also define the field strength $F_{\mu\nu}$ of the Weyl vector $\omega_\mu$ according to
\begin{equation}  \label{W}
\tilde{F}_{\mu\nu} = \tilde{\nabla}_{\mu} \omega_{\nu} - \tilde{\nabla}
_{\nu} \omega_{\mu} = \nabla _{\mu}\omega _\nu-\nabla _\nu \omega _\mu.
\end{equation}

 The tensor curvatures and the curvature scalar of Weyl geometry can be obtained by using definitions
similar to the ones used in the Riemannian geometry, but with the Levi-Civita connection $\Gamma $
replaced with the Weyl connection $\tilde\Gamma$. Thus, with the help of the explicit expression of the Weyl connection $\tilde\Gamma$ we obtain
\begin{align}
\tilde R^\lambda_{\mu\nu\sigma}&=\partial_\nu
\tilde\Gamma^\lambda_{\mu\sigma} -\partial_\sigma
\tilde\Gamma^\lambda_{\mu\nu}
+\tilde\Gamma^\lambda_{\nu\rho}\,\tilde\Gamma_{\mu\sigma}^\rho
-\tilde\Gamma_{\sigma\rho}^\lambda\,\tilde\Gamma_{\mu\nu}^\rho,  \notag \\
\tilde R_{\mu\nu}&=\tilde R^\lambda_{\mu\lambda\sigma}, \;\;\;\tilde
R=g^{\mu\sigma}\,\tilde R_{\mu\sigma}.
\end{align}

In terms of the Riemannian quantities, and of the Weyl vector,  one finds
\begin{align}  \label{tRmunu}
\tilde R_{\mu\nu}=R_{\mu\nu} &+\frac {1}{2}\alpha \left(\nabla_\mu \omega
_\nu-3\,\nabla_\nu \omega _\mu - g_{\mu\nu}\,\nabla_\lambda \omega
^\lambda\right)  \notag \\
&+\frac{1}{2} \alpha ^2 (\omega _\mu \omega _\nu -g_{\mu\nu}\,\omega
_\lambda \omega ^\lambda),
\end{align}
\begin{align}
\tilde R_{\mu\nu}&-\tilde R_{\nu\mu}=2 \alpha F_{\mu\nu},
\end{align}
and
\begin{align}\label{tR}
\tilde R&= R-3 \,\alpha\,\nabla_\lambda\omega ^\lambda-\frac{3
}{2}\alpha ^2 \omega _\lambda \omega ^\lambda,
\end{align}
where $R_{\mu \nu}$ and $R$ are the Ricci curvature tensor, and the Ricci
scalar curvature of the Riemannian geometry. 

The right hand sides of the
above equations are defined in the Riemannian geometry.
It should also be noted that the Weyl scalar $\tilde R$ transforms covariantly under
the conformal transformations, so that $\hat{\tilde R}=(1/\Sigma^n)\,\tilde R$.

\subsection{The boundary term in general relativity}

We consider now the Einstein-Hilbert action of standard general relativity, which is defined in the Riemannian geometry. The Einstein-Hilbert action is
constructed solely from the Ricci scalar, and the matter action $S_m$, and it is given by \cite{LaLi}
\begin{equation}\label{act1}
S_{g}=-\frac{1}{2\kappa^2}\int_{\Omega }{R(g)\sqrt{-g}d^{4}x}+S_m,
\end{equation}
where we have denoted $\kappa^{2}=8\pi G/c^4$. The variation of the Ricci
scalar with respect to the metric is obtained in a straightforward way as
\begin{align}
\delta \left( R\sqrt{-g}\right) &=\delta \left( R_{\mu \nu }g^{\mu \nu }
\sqrt{-g}\right)  \notag \\
&=\left( R_{\mu \nu }-\frac{1}{2}g_{\mu \nu }R\right) \sqrt{-g}\delta
g^{\mu \nu }
+g^{\mu \nu }\delta R_{\mu \nu }\sqrt{-g}. \nonumber\\
\end{align}

For the variation of $R_{\mu \nu }$ we find \cite{LaLi} 
\begin{equation}
\delta R_{\mu \nu }=\nabla _{\lambda }\delta \Gamma _{\mu \nu }^{\lambda
}-\nabla _{\nu }\delta \Gamma _{\mu \sigma }^{\sigma },
\end{equation}
which gives immediately
\begin{equation}
g^{\mu \nu }\delta R_{\mu \nu }=g^{\mu \nu }\nabla _{\lambda }\delta \Gamma
_{\mu \nu }^{\lambda }-g^{\mu \lambda }\nabla _{\lambda }\delta \Gamma _{\mu
\sigma }^{\sigma }.
\end{equation}
Therefore, the variation of the Einstein-Hilbert action is given in a general form by
%\begin{widetext}
\begin{align}\label{gravact}
	\delta S_{g}=&-\frac{1}{2\kappa^2}\int_\Omega \Big( G_{\mu \nu }\delta g^{\mu \nu }+g^{\mu
		\nu }\nabla _{\lambda }\delta \Gamma _{\mu \nu }^{\lambda }\nonumber\\&-g^{\mu \lambda
	}\nabla _{\lambda }\delta \Gamma _{\mu \sigma }^{\sigma }-\kappa^2 T_{\mu\nu}\delta g^{\mu\nu}\Big)\sqrt{-g}d^{4}x,
\end{align}
%\end{widetext}
where we have introduced the Einstein tensor as
$$G_{\mu \nu }=R_{\mu \nu }-\frac12Rg_{\mu \nu },$$
and defined the matter energy-momentum tensor as
$$T_{\mu\nu}=-\frac{2}{\sqrt{-g}}\frac{\delta(\sqrt{-g}L_m)}{\delta g^{\mu\nu}}.$$

The variation of the Christoffel symbol with respect to the metric can be obtained in a
covariant form as
\begin{equation}
\delta \Gamma _{\mu \nu }^{\lambda }=\frac{1}{2}g^{\sigma \lambda }\left[
\nabla _{\nu }\delta g_{\mu \sigma }+\nabla _{\mu }\delta g_{\nu \sigma
}-\nabla _{\sigma }\delta g_{\mu \nu }\right] .  \label{WC}
\end{equation}

In standard general relativity, based on the Riemannian geometric formulation, the
boundary term $g^{\mu \nu }\delta R_{\mu \nu }\sqrt{-g}$ is canceled by
 extending the boundary to infinity, by using the Gauss's theorem, and assuming that the variations of the metric tensor vanish on $\partial \Omega
$,
\begin{equation}
\int_{\Omega }g^{\mu \nu }\delta R_{\mu \nu }\sqrt{-g}d^{4}x=\int_{\Omega
}\nabla _{\lambda }A^{\lambda }\sqrt{-g}d^{4}x=\int_{\partial \Omega
}A^{\lambda }dS_{\lambda },
\end{equation}%
where we have denoted $A^\lambda=g^{\mu\nu}\delta\Gamma^\lambda_{\mu\nu}-g^{\mu\lambda}\delta\Gamma^\sigma_{\mu\sigma}$. Moreover,  $dS_{\lambda }$ represents the element of integration on the hypersurface $\partial \Omega$ surrounding the four-volume element $d\Omega $.

\section{Gravitational action, and field equations with Weylian boundary terms}\label{derfield}

We will generalize now the Einstein-Hilbert action (\ref{act1}) by following the approach introduced in \cite{Harko},  by explicitly maintaining the boundary terms.  Moreover, we also introduce  a  Weylian geometric framework, which we assume is realized on the boundary $\partial \Omega$ of the space-time.  Thus, we disregard the assumption of the existence of a metric geometry 
 on $\partial \Omega$. This implies that near and on the boundary the covariant derivative of the metric tensor is given 
by Eq.~(\ref{nm}), leading to the replacement $\nabla \rightarrow \tilde{\nabla}$. Moreover, 
the Christoffel connection, as well as its variation, must be replaced by the
Weyl connection, $\Gamma \rightarrow \tilde{\Gamma}$, and by $\delta \Gamma
\rightarrow \delta \tilde{\Gamma}$. Therefore, based on the above assumptions, in the calculation of the variation of the action,  we replace the Riemannian boundary conditions by their Weylian counterparts. The consideration of the Weylian boundary condition will lead to a modification of the standard gravitational field equations of general relativity.

We reconsider now the variation of the Einstein-Hilbert action (\ref{act1}) by assuming the presence of non-vanishing boundary terms in the integral. The variation of the action with respect to the metric leads to the general expression \cite{Harko}
\bea\label{25}
\delta S_g&=&-\frac{1}{2\kappa ^2}\int_\Omega\left(R_{\mu \nu}-\frac{1}{2}Rg_{\mu \nu}-\kappa^2 T_{\mu\nu}\right)\delta g^{\mu\nu}\sqrt{-g}d^4x\nonumber\\
&&-\frac{1}{2\kappa ^2}\int_\Omega {\left(g^{\mu
		\nu }\nabla _{\lambda }\delta \Gamma _{\mu \nu }^{\lambda }-g^{\mu \lambda
	}\nabla _{\lambda }\delta \Gamma _{\mu \sigma }^{\sigma }\right)\sqrt{-g}d^4x}.\nonumber\\
\eea

We consider now the action (\ref{25}) for a particular choice of the boundary term. More exactly, we assume that on the boundary the Riemannian connection, given by the Christoffel symbols, as well as their variations with respect to the metric tensor, are replaced by their analogues defined in the Weyl geometric framework. Since the choice of the boundary terms is arbitrary, such a redefinition is always possible. Hence, we propose a gravitational theory in which in the expressions of the variations of the Christoffel symbols, the Riemannian covariant derivative is replaced with the Weylian one. Accordingly, we obtain the following Weyl geometric type extension of the Einstein-Hilbert action \cite{Harko}
\begin{align}\label{25.1}
\delta S_g=&-\frac{1}{2\kappa ^2}\int_\Omega\left(R_{\mu \nu}-\frac{1}{2}Rg_{\mu \nu}-\kappa^2 T_{\mu\nu}\right)\delta g^{\mu\nu}\sqrt{-g}d^4x\nonumber\\
&-\frac{1}{2\kappa ^2}\int_\Omega {\left(g^{\mu
		\nu }\tilde\nabla _{\lambda }\delta \tilde\Gamma _{\mu \nu }^{\lambda }-g^{\mu \lambda
	}\tilde\nabla _{\lambda }\delta \tilde\Gamma _{\mu \sigma }^{\sigma }\right)\sqrt{-g}d^4x}.\nonumber\\
\end{align}

\subsection{Variation of the Weylian boundary terms}

The details of the variation of the variation with respect to the metric of the Weylian boundary terms are presented in \cite{Harko}, but for the sake of completeness we briefly review them here. The condition $g_{\mu \nu }g^{\nu \sigma }=\delta
_{\mu }^{\sigma }$, immediately gives
$
\delta g_{\mu \lambda }=-g_{\mu \nu }g_{\lambda \sigma }\delta g^{\nu \sigma
},
$
while from the definition of the Weyl connection we obtain
\begin{equation}
\tilde{\nabla}_{\lambda }g^{\mu \nu }=\alpha \omega _{\lambda }g^{\mu \nu }.
\end{equation}

 Furthermore, we introduce the important assumption that the boundary Weyl vector is small, and that the variation operator $\delta$ commutes with the covariant derivative $\tilde{\nabla}_{\sigma }$.  Hence for the Weyl covariant derivative of the variation of the metric tensor we obtain
\begin{equation}\label{29}
\tilde{\nabla}_{\sigma }\delta g_{\mu \nu }=-\alpha \omega _{\sigma }\delta
g_{\mu \nu }=\alpha \omega _{\sigma }g_{\gamma \mu }g_{\rho \nu }\delta
g^{\gamma \rho }.
\end{equation}
\newline
%Hence, we obtain first
%\begin{equation}
%\tilde{\nabla}_{\lambda }g^{\mu \nu }-\delta _{\lambda }^{\nu }\tilde{\nabla}%
%_{\sigma }g^{\mu \sigma }=\alpha \left( \omega _{\lambda }g^{\mu \nu
%}-\delta _{\lambda }^{\nu }\omega ^{\mu }\right) .
%\end{equation}

For the variation of the boundary Weyl connection, given by
\begin{equation}
\delta \tilde{\Gamma}_{\mu \nu }^{\lambda }=\frac{1}{2}g^{\sigma \lambda }%
\left[ \tilde{\nabla}_{\nu }\delta g_{\mu \sigma }+\tilde{\nabla}_{\mu
}\delta g_{\nu \sigma }-\tilde{\nabla}_{\sigma }\delta g_{\mu \nu }\right] ,
\label{WC1}
\end{equation}%
we find, with the use of Eq. (\ref{29}), the expression
\begin{equation}
\delta \tilde{\Gamma}_{\mu \nu }^{\lambda }=\frac{\alpha }{2}\left( \omega
_{\nu }g_{\mu \rho }\delta _{\gamma }^{\lambda }+\omega _{\mu }g_{\nu \gamma
}\delta _{\rho }^{\lambda }-\omega ^{\lambda }g_{\mu \gamma }g_{\nu \rho
}\right) \delta g^{\rho \gamma }.
\end{equation}

Therefore, we obtain
\begin{align}
\tilde{\nabla}_{\lambda }\delta \tilde{\Gamma}_{\mu \nu
}^{\lambda } &=\frac{\alpha }{2}\Bigg[\left( \tilde{\nabla}_{\gamma }\omega
_{\nu }\right) g_{\mu \rho }+\left( \tilde{\nabla}_{\rho }\omega _{\mu
}\right) g_{\nu \gamma }  \notag \\
&-\left( \tilde{\nabla}_{\lambda }\omega ^{\lambda }\right)
g_{\mu \gamma }g_{\nu \rho }+\alpha g_{\mu \gamma }g_{\nu \rho }\omega
^{\lambda }\omega _{\lambda }\Bigg]\delta g^{\rho \gamma }.
\end{align}%

Thus, the first Weylian boundary term is given by \cite{Harko}
\begin{eqnarray}
g^{\mu \nu }\tilde{\nabla}_{\lambda }\delta \tilde{\Gamma}_{\mu \nu
}^{\lambda }&=&\frac{\alpha }{2}\Big[\tilde{\nabla}_{\gamma }\omega _{\rho }
+\tilde{\nabla}_{\rho }\omega _{\gamma } - g_{\gamma \rho}\tilde{\nabla}%
_{\lambda }\omega ^{\lambda }  \notag \\
&&+\alpha g_{\gamma \rho }\omega ^{\lambda }\omega _{\lambda }\Big] \delta
g^{\rho \gamma }.
\end{eqnarray}

 By using the alternative form of the Weylian variation of  the connection 
\begin{equation}
\delta \tilde{\Gamma}_{\mu \sigma }^{\sigma }=\frac{\alpha }{2}\left( \omega
_{\gamma }g_{\mu \rho }+\omega _{\mu }g_{\gamma \rho }-\omega _{\rho }g_{\mu
\gamma }\right) \delta g^{\rho \gamma },
\end{equation}
we easily find for the second Weylian boundary term the expressions
\begin{equation}
\tilde{\nabla}_{\lambda }\delta \tilde{\Gamma}_{\mu \sigma }^{\sigma }=\frac{
\alpha }{2}[(\tilde{\nabla}_{\lambda }\omega _{\rho })g_{\mu \gamma }+(%
\tilde{\nabla}_{\lambda }\omega _{\mu })g_{\gamma \rho }-(\tilde{\nabla}%
_{\lambda }\omega _{\rho })g_{\mu \gamma }]\delta g^{\rho \gamma },
\end{equation}
and
\begin{align}
g^{\mu \lambda }\tilde{\nabla}_{\lambda }\delta \tilde{\Gamma}_{\mu \sigma
}^{\sigma } &=\frac{\alpha }{2}\left[ \Big(\tilde{\nabla}_{\rho }\omega
_{\gamma }\right) +g^{\mu \lambda }\left( \tilde{\nabla}_{\lambda }\omega
_{\mu }\right) g_{\gamma \rho }  \notag \\
&-\left( \tilde{\nabla}_{\gamma }\omega _{\rho }\right) \Big]\delta g^{\rho
\gamma },
\end{align}%
respectively. Thus the total variation of the Weylian boundary
terms can be obtained as \cite{Harko}
%\begin{widetext}
\begin{align}\label{40}
&g^{\mu \nu }\tilde{\nabla}_{\lambda }\delta \tilde{\Gamma}%
_{\mu \nu }^{\lambda }-g^{\mu \lambda }\tilde{\nabla}_{\lambda }\delta
\tilde{\Gamma}_{\mu \sigma }^{\sigma }=\frac{\alpha }{2}\delta g^{\rho \gamma }\times \nonumber\\
& \Big[\Big(2\tilde{%
\nabla}_{\gamma }\omega _{\rho }-g_{\gamma \rho }\tilde{\nabla}_{\lambda
}\omega ^{\lambda }
-g^{\mu \lambda }g_{\gamma \rho }\tilde{\nabla}_{\lambda
}\omega _{\mu }\Big)
+\alpha \omega ^{\lambda }\omega
_{\lambda }g_{\gamma \rho }\Big].\nonumber\\
\end{align}
%\end{widetext}

\subsection{Einstein-Hilbert action and gravitational field equations with Weyl type boundary terms}

Now,  by using Eq.~(\ref{40}), we  can write down the variation with respect to the metric of the Einstein-Hilbert gravitational action
 in the presence of Weyl type boundary terms, which takes the form \cite{Harko}
%\begin{widetext}
\begin{align}
\delta \tilde{S}_{g} = &-\frac{1}{2\kappa^2} \int_{\Omega }\sqrt{-g}%
d^{4}x\Bigg[ G_{\mu \nu }+\frac{\alpha ^2}{2}g_{\mu \nu}\omega_\lambda \omega ^\lambda \nonumber\\
&+\frac{%
	\alpha }{2}\left( \tilde{\nabla}_{\mu }\omega _{\nu }+\tilde{\nabla}_{\nu }\omega _{\mu }-g_{\mu \nu }\tilde{%
\nabla}_{\lambda }\omega ^{\lambda }-g^{ \sigma \lambda }g_{\mu \nu }\tilde{%
\nabla}_{\lambda }\omega _{\sigma }\right)\nonumber\\&-\kappa^2 T_{\mu\nu}
\Bigg] \delta g^{\mu \nu }.
\end{align}%
%\end{widetext}

By replacing the expressions of the Weyl covariant derivatives as given by Eqs.~(\ref{cov1}) and (\ref{cov2}), we finally obtain the gravitational field equations in the presence of Weylian boundary terms,  as
\begin{align}\label{feqa}
G_{\mu \nu }&+\frac12\alpha \left(\nabla _\mu \omega _\nu+\nabla _\nu \omega _\mu-2g_{\mu \nu}\nabla _\lambda \omega ^\lambda\right)\nonumber\\&-\alpha ^2\left(\omega _\mu \omega _\nu+\frac{1}{2}\omega ^2 g_{\mu \nu}\right)=\kappa^2T_{\mu\nu}, \nonumber\\
\end{align}
where $\omega ^2\equiv\omega _\lambda \omega ^\lambda$.
The field equations (\ref{feqa}) are defined in the standard metric Riemann geometry, with the covariant derivatives defined with the help of the Levi-Civita connection. The Weyl type boundary terms do add to the field equations several new term, defined in a Weyl geometric framework, and which depend on the Weyl gauge vector, and of its covariant derivatives.

By taking the covariant derivative of equation \eqref{feqa}, one can obtain the conservation equation of the matter sector as
\begin{align}\label{conseq}
\kappa^2\nabla^\nu T_{\mu\nu}&=\frac12\alpha(\omega^\alpha R_{\mu\alpha}+\Box \omega_\mu-\nabla_\mu\nabla_\alpha\omega^\alpha)\nonumber\\
&-\alpha^2(\omega_\mu\nabla_\alpha\omega^\alpha+\omega^\alpha\nabla_\alpha\omega_\mu+\omega^\alpha\nabla_\mu\omega_\alpha)
\end{align}
It should be noted that since the field equation \eqref{feqa} does not directly obtained from an action principle the matter energy-momentum tensor is not conserved in this model. As one can see from equation \eqref{conseq} the non-conservation of the energy-momentum tensor is fully dependent on the Weyl vector which comes from the boundary term.

 \section{Cosmological applications}\label{cosm}
 
In this Section we will consider the cosmological implications of the gravitational field equations in the presence of a Weyl boundary. Let us assume that the Universe can be described by a flat, homogeneous and isotropic FLRW metric, with line element
\begin{align}\label{metric}
	ds^2 = dt^2-a^2(t)\left(dx^2+dy^2+dz^2\right),
\end{align}
where $a(t)$ is the scale factor, and $t$ denotes the cosmological time. We also introduce the Hubble parameter, defined as $H=\dot{a}/a$, where a dot denotes the derivative with respect to the cosmological time $t$. 

The matter content of the Universe is assumed to be a perfect fluid, with energy momentum tensor of the form
\begin{align}
	T^\mu_\nu=\mathrm{diag}(\rho,-p,-p,-p),
\end{align}
where $\rho$ and $p$ are the energy density and thermodynamical pressure of the cosmological matter source.

\subsection{The generalized Friedmann equations}

For an isotropic and homogeneous Universe with a metric of the form \eqref{metric}, the gauge Weyl vector field must have the form
\begin{align}
	\omega_\mu=(\omega_0(t),0,0,0),
\end{align}
where $\omega_0(t)$ is an arbitrary function of time.
With the above assumptions, the cosmological field equations in the presence of a Weyl type boundary, the generalized Friedmann equations,  can be obtained as
\be\label{Fr1}
	3H^2=\kappa^2\rho+\frac32\alpha\omega_0(2H+\alpha\omega_0)=\kappa^2 (\rho+\rho_{eff}),
\ee
and
\bea\label{Fr2}
	2\dot{H}+3H^2&=&-\kappa^2p+\frac12\alpha\omega_0(4H+\alpha\omega_0)+\alpha\dot\omega_0\nonumber\\
&=&-\kappa ^2 (p+p_{eff}),
\eea
where we have denoted the effective energy density and pressure as
\be\label{rhoeff}
\rho_{eff}=\frac32\alpha\omega_0(2H+\alpha\omega_0),
\ee
and
\be\label{peff}
p_{eff}=-\frac12\alpha\omega_0(4H+\alpha\omega_0)-\alpha\dot\omega_0,
\ee
respectively.

Also, the non-conservation equation of the matter energy-momentum tensor can be obtained as
\begin{align}\label{Fr3}
	\dot{\rho}+3H(\rho+p) = -\frac{3\alpha}{\kappa^2}\omega_0\left[H(\alpha\omega_0+H)+\alpha\dot\omega_0+\dot{H}\right].
\end{align}

 From the generalized Friedmann equations one can obtain the time evolution of $H$ as
\be
\dot{H}=-\frac{1}{2}\kappa ^2(\rho+p)-\frac{1}{2}\alpha \omega _0\left(H+\alpha \omega _0\right)+\frac{1}{2}\alpha \dot{\omega}_0.
\ee

The deceleration parameter $q$ is defined according to
\be
q=\frac{d}{dt}\frac{1}{H}-1=-\frac{\dot{H}}{H^2}-1,
\ee
which can be obtained as
\be
q=\frac{\kappa ^2(\rho+3p)-6\alpha \omega _0H-3\alpha \dot{\omega}_0}{2\kappa ^2\rho+3\alpha\omega _0\left(2H+\alpha \omega _0\right)}.
\ee

In this work, we will concentrate on the late time behavior of the Universe. Consequently, we will assume that the matter content of the Universe can be described by a pressure-less fluid, with $p=0$ and $\rho = \rho_m$, where $\rho_m$ is the baryonic matter density.

\subsubsection{Dimensionless representation}

Now, let us perform the following transformation to a set of dimensionless variables $\left(\tau, \bar\rho , \bar\omega_0\right)$
\begin{align}
	\tau = H_0 t,\quad H=H_0 h, \quad \bar\rho_m = \frac{\kappa^2\rho_m}{3H_0^2},\quad \omega_0 = H_0 \bar\omega_0,
\end{align}
and then transform to the redshift coordinate $z$, defined as
\begin{align}
	1+z=\frac1a.
\end{align}

Since the Weyl vector comes from the boundary, there is no equation of motion for it. 

In this paper, in order to fix the dynamics of the Weyl vector field, we will assume that the effective dark energy density and pressure, defined in Eqs.~\eqref{rhoeff} and \eqref{peff}, satisfy the following equation of state
\begin{align}\label{eoseff}
	p_{eff}=\gamma(z)\rho_{eff},
\end{align}
where $\gamma (z)$ is the equation of state parameter, for which we adopt the Barboza-Alcaniz parametrization \cite{barbozaalcaniz}, defined as
\begin{align}
	\gamma(z) = \gamma_0 +\gamma_a \frac{z(1+z)}{1+z^2},
\end{align}
where $\gamma _0$ and $\gamma _a$ are constants.
 This parametrization has the property that behaves correctly for both small and large values of $z$, and it is also finite at the future boundary $z=-1$.
 
Therefore, using Eqs.~\eqref{Fr1}, \eqref{Fr2} and \eqref{Fr3}, one can  obtain the Hubble parameter as
\begin{align}\label{hubble}
	h(z)= \omega_1+\sqrt{3\omega_1^2+\bar\rho_m},
\end{align}
where we have defined $$\omega_1=\frac\alpha2\bar\omega_0,$$ and the functions $\bar\rho_m(z)$ and $\bar{\omega}_1(z)$ are given as the solutions of the following system of differential equations
\begin{align}\label{ode1}
	\bar\rho_m^\prime(z)& = \frac{1}{(1+z)(\omega_2^2-\omega_1^2)}\Big[(3\omega_2^2-2\omega_1^2-\omega_1\omega_2)\bar\rho_m\nonumber\\&-6(1+3\gamma)(\omega_2^2-2\omega_1^2-\omega_1\omega_2)\omega_1^2\Big],
\end{align}
\begin{align}\label{ode2}
	\bar{\omega}_1^\prime(z) = \frac{\omega_1}{(1+z)(\omega_2^2-\omega_1^2)}&\Big[(1+\gamma)\omega_2^2+(1+2\gamma)\bar\rho_m\nonumber\\&+(1+3\gamma)\omega_1\omega_2\Big], 
\end{align}
where we have denoted
$$\omega_2 = \sqrt{3\omega_1^2+\bar\rho_m}.$$

One should note that with the definitions above, the parameter $\alpha$ is completely absorbed in $\bar{\omega}_0$. As a result, we keep the variable $\omega_1$ as a dynamical variable. Also, using the fact that $h(z=0)=1$, we obtain the constraint
\begin{align}\label{omega1}
	\omega_1(z=0) =  \frac12(-1+\sqrt{3-2\Omega_{m0}}),
\end{align}
where $\Omega_{m0}$ is the current value of the matter density parameter.

\subsection{Statistical analysis}

From Eqs.~\eqref{hubble},\eqref{ode1} and \eqref{ode2} one can see that the free parameters of the model are $H_0$, $\Omega_{m0}$, $\gamma_0$ and $\gamma_a$. To find the best fit values of these parameters, we will perform the Likelihood analysis with the following datasets.
\begin{itemize}
\item\textbf{Cosmic Chronometers:}
We use the Hubble parameter $H(z)$ measurements from $31$ data points derived from the relative ages of massive, early-time, passively
evolving galaxies, known as cosmic chronometers (CC) \cite{CCdata}. These are assumed to be independent. 
\item\textbf{SNe Ia:}
we use the SNe Ia distance moduli measurements from the Pantheon$^+$ sample \cite{PANdata}, which consists of $1701$ light curves of $1550$ distinct SNe Ia, ranging in the redshift interval $z\in[0.001, 2.26]$. 
From the theoretical part, the distance module $\mu$ can be obtained from the relation
\begin{align}
	\mu = m_b - \mathcal{M} = 25 + 5\log_{10}d_L(z),
\end{align}
where $\mathcal{M}$ is the absolute magnitude and the luminosity distance $d_L$ is defined as
\begin{align}
	d_L = (1+z)\int_0^z \frac{c}{H(z^\prime)}dz^\prime.
\end{align}
In this paper we will use the dataset without SH0ES calibration and the absolute magnitude $\mathcal{M}$ will be derived from the fitting \cite{noshoes}.
\item\textbf{BAO:}
We use recent data from DESI DR2 \cite{bao}, including BGS, LRGs, ELGs, QSOs and Lyman-$\alpha$ forests.
The data includes the anisotropic BAO measurements of $D_M /r_d$ , $D_H /r_d$ where $D_M$ and $D_H$ are comoving angular diameter and Hubble distances defined as
\begin{align}
	D_A = \frac{1}{1+z}\int_0^z\frac{c}{H(z^\prime)}dz^\prime,
\end{align}
and
\begin{align}
	D_M = (1+z)D_A,
\end{align}
and the isotropic BAO measurements of $D_V /r_d$ with $D_V$ is the spherically averaged volume distance defined as
\begin{align}
	D_V = \left((1+z)^2D_A(z)^2\frac{cz}{H(z)}\right)^\frac13.
\end{align}
In this paper, we will obtain the value of the sound horizon $r_d$ from the fitting.
\end{itemize}
The likelihood function can be defined as
\begin{align}
	L=L_0e^{-\chi^2/2},
\end{align}
where $L_0$ is the normalization constant. The loss functions $\chi^2$ for the cosmic chronometer and Pantheon$^+$ and BAO data points are defined as
\begin{align}
	\chi_{CC}^2=\sum_i\left(\frac{{H}_{\text{obs},i}-{H}_{\text{th},i}}{\sigma_i}\right)^2,
\end{align}
\begin{align}
	\chi^2_{\text{Pantheon$^+$}} = \left[\vec{\mu}_{\text{obs}} - \vec{\mu}_{\text{th}}\right]^T 
	C^{-1} 
	\left[\vec{\mu}_{\text{obs}} - \vec{\mu}_{\text{th}}\right],
\end{align}
and 
\begin{align}
		\chi^2_{\text{BAO}} =\Delta D_a^T 
	C_a^{-1} 
\Delta D_a,
\end{align}
with
\begin{align}
	 \Delta D_a = \left(\frac{D_{a}}{r_d}\right)_\text{obs} - \frac{D_{a,\text{th}}}{r_d},
\end{align}
where $a$ stands for $V$, $M$ and $H$. In the above expressions $i$ counts data points, $``obs"$ are the observational values $``th"$ are the theoretical values obtained from the model, and $\sigma_i$ are the errors associated with the $i$th data obtained from observations. Also, $C$'s are the covariance matrix associated with Pantheon$^+$ and BAO datapoints.

By maximizing the likelihood function, the best fit values of the parameters $H_0$, $\Omega_{m0}$, $\gamma_0$, $\gamma_a$, $\mathcal{M}$ and $r_d$ at $1\sigma$ confidence level, can be obtained. In table \eqref{bestfit}, we have presented the result of four different choices of the datasets, including CC,  Pantheon$^+$,  CC+ BAO,  and CC+ Pantheon$^+$+ BAO, respectively.

\pagebreak
%\begin{sidewaystable}
%\setlength{\tabcolsep}{3pt}
%\renewcommand{\arraystretch}{1.5}
\begin{widetext}
%\centering
\begin{table}[htbp]
\resizebox{18cm}{!}{%
\begin{tabular}{|c||c|c|c|c||c|c|c|c|}
		\hline
		\multirow{2}{*}{\textbf{Parameters}}& \multicolumn{4}{c||}{\textbf{$\Lambda$CDM}} &\multicolumn{4}{c|}{\textbf{Weyl Boundary}} \\ \cline{2-9}
		 & CC & CC+Pantheon$^+$ & CC+BAO & CC+Pantheon$^+$+BAO & CC& CC+Pantheon$^+$ & CC+BAO & CC+Pantheon$^+$+BAO \\ \hline\hline
		\rule{0pt}{14pt} $\mathrm{H_{0}}$ & $67.746_{-3.086}^{+2.986}$ & $67.178_{-1.710}^{+1.832}$ & $69.163_{-1.543}^{+1.551}$ & $68.653_{-1.631}^{+1.625}$ & $65.282_{-3.943}^{+3.967}$ & $67.305_{-1.603}^{+1.566}$ & $66.090_{-1.731}^{+1.645}$ & $67.308_{-1.666}^{+1.608}$ \\ \hline
		\rule{0pt}{14pt} $\mathrm{\Omega_{m0}}$ & $0.331_{-0.060}^{+0.058}$ & $0.343_{-0.017}^{+0.016}$ & $0.298_{-0.007}^{+0.006}$& $0.307_{-0.008}^{+0.008}$ & $0.186_{-0.116}^{+0.117}$ & $0.136_{-0.047}^{+0.043}$ & $0.144_{-0.028}^{+0.027}$ & $0.111_{-0.017}^{+0.017}$ \\ \hline
		\rule{0pt}{14pt} $\mathrm{\gamma_0}$ & -- & -- & -- & --& $-0.609_{-0.203}^{+0.204}$ & $-0.671_{-0.077}^{+0.074}$ & $-0.554_{-0.046}^{+0.042}$ & $-0.609_{-0.031}^{+0.030}$ \\ \hline
		\rule{0pt}{14pt} $\mathrm{\gamma_a}$ & -- & -- & -- & --& $-0.524_{-0.254}^{+0.262}$ & $-0.653_{-0.100}^{+0.100}$ & $-0.619_{-0.009}^{+0.009}$ & $-0.597_{-0.082}^{+0.073}$ \\ \hline
		\rule{0pt}{14pt} $\mathcal{M}$ & -- & $-19.435_{-0.054}^{+0.055}$ & --& $-19.400_{-0.051}^{+0.051}$ & -- & $-19.422_{-0.053}^{+0.049}$ & --&  $-19.416_{-0.051}^{+0.050}$ \\ \hline
		\rule{0pt}{14pt} $\mathrm{r_d}$ & -- & -- & $146.789_{-3.222}^{+3.200}$ &  $146.906_{-3.367}^{+3.382}$ & -- & -- & $147.403_{-3.123}^{+3.162}$ & $147.293_{-3.365}^{+3.498}$ \\ \hline
	\end{tabular}%
}
\caption{The best fit values together with their 1$\sigma$ errors for the Weyl boundary gravity. The $\Lambda$CDM values are also reported for the sake of comparison.}\label{bestfit}
%\end{center}
\end{table}
\end{widetext}
%\end{sidewaystable}

\pagebreak
As one can see from Table~\ref{bestfit}, the present day values of the Hubble parameter, as predicted by both models, are very close to each other. However, a significant difference does appear in the matter density parameter, with the $\Lambda$CDM model predicting the existence of a much larger amount of matter. Both coefficients $\gamma _0$ and $\gamma _a$ in the parameter of the dark energy equation of state have negative values.

\begin{figure*}
	\includegraphics[scale=0.4]{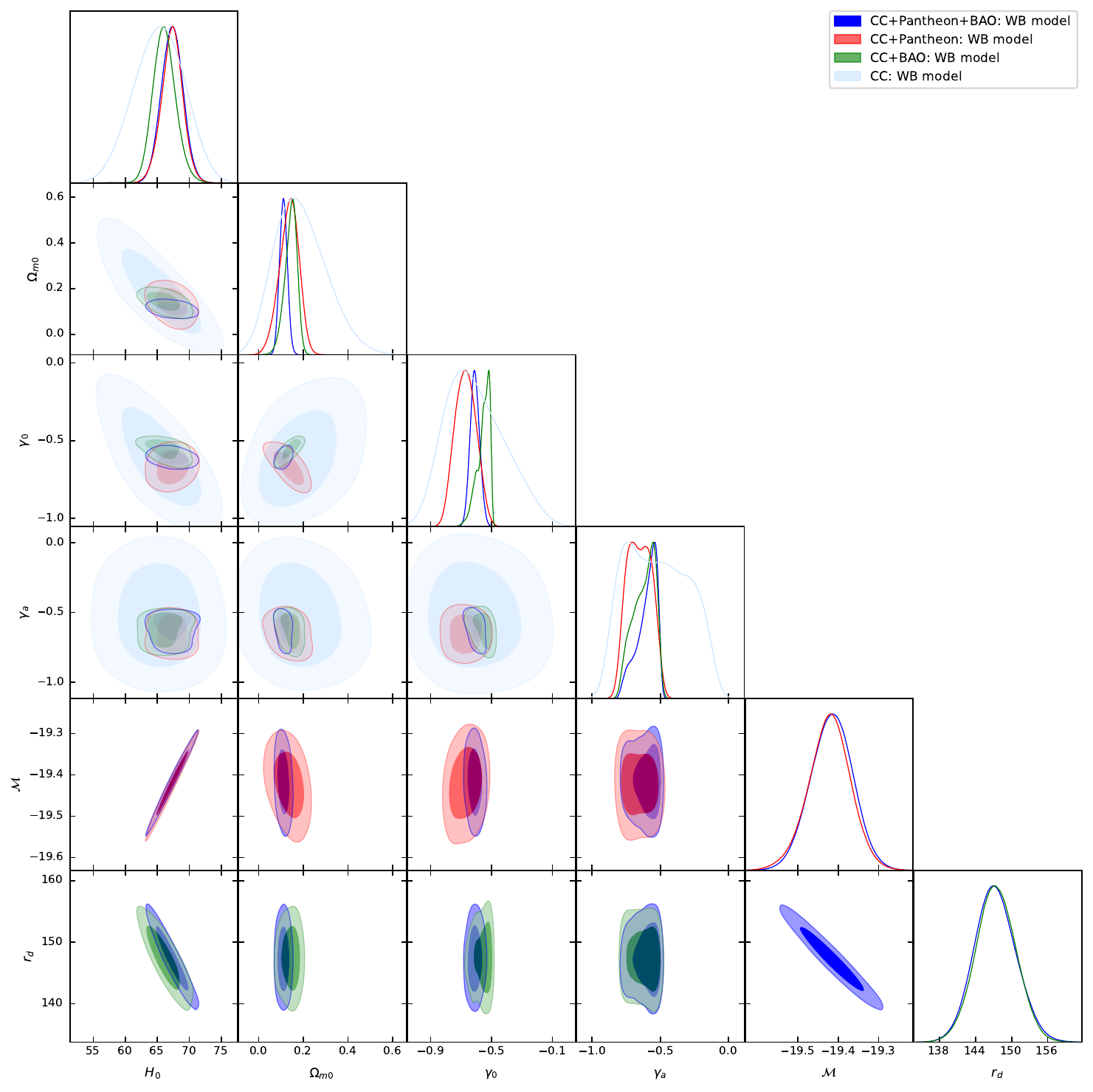}\includegraphics[scale=0.4]{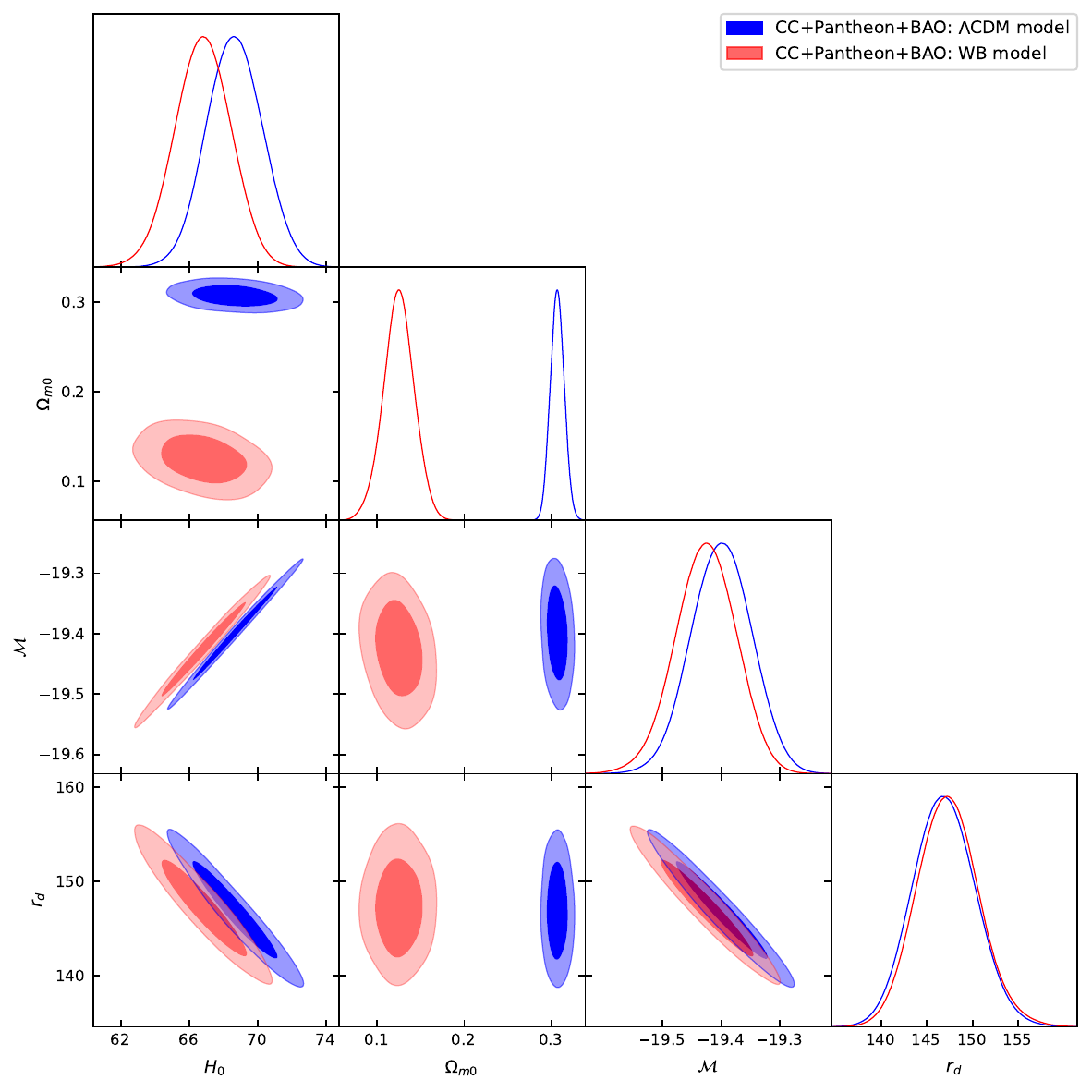}
	\caption{The corner plot for the values of the parameters $H_0$, $\Omega_{m0}$, $\gamma_0$ and $\gamma_a$ with their $1\sigma$ and $2\sigma$ confidence levels for the Weyl Boundary gravity model. \label{cornerplot}}
\end{figure*}

The corner plot for the values of the parameters $H_0$, $\Omega_{m0}$, $\gamma_0$, $\gamma_a$, $\mathcal{M}$ and $r_d$ with their $1\sigma$ and $2\sigma$ confidence levels is shown in Fig.~\ref{cornerplot}. In this figure we have also plotted the corner plot of the parameters $H_0$, $\Omega_{m0}$, $\mathcal{M}$ and $r_d$ with their $1\sigma$ and $2\sigma$ confidence levels for the Weyl Boundary and $\Lambda$CDM models for comparison.

The differences between the Weyl Boundary geometric gravity cosmological model and the $\Lambda$CDM standard paradigm can also be seen through the values of the reduced chi-squared function for the Hubble parameter, which are shown in Table~\ref{table1}.

\begin{table}[h!]
	\centering
	\begin{tabular}{|c||c|c|c|c|}
		\hline
		\multirow{2}{*}{\textbf{Model}}&\multicolumn{4}{c|}{\textbf{$\chi_{red}^2$}}\\ \cline{2-5}
		& \small{CC} & CC, Pnt$^+$ & CC, BAO & CC, Pnt$^+$, BAO  \\\hline\hline
		\textbf{$\Lambda$CDM} & 0.569 & 1.045 & 0.668 & 1.048 \\\hline
		\textbf{WB}  & 0.624 & 1.046 & 0.666 &  1.045 \\\hline
	\end{tabular}
	\caption{The reduced-$\chi^2$ for the Weyl Boundary (WB) gravity and the $\Lambda CDM$ model for various choices of datasets.}\label{table1}
\end{table}

As one can see from Table~\ref{table1}, the values of $\chi_{red}^2$ basically coincide for both the Weyl boundary and $\Lambda$CDM cosmological models, indicating that both models can describe the considered set of observational data at a similar level of precision.  

\begin{figure*}
	\includegraphics[scale=0.5]{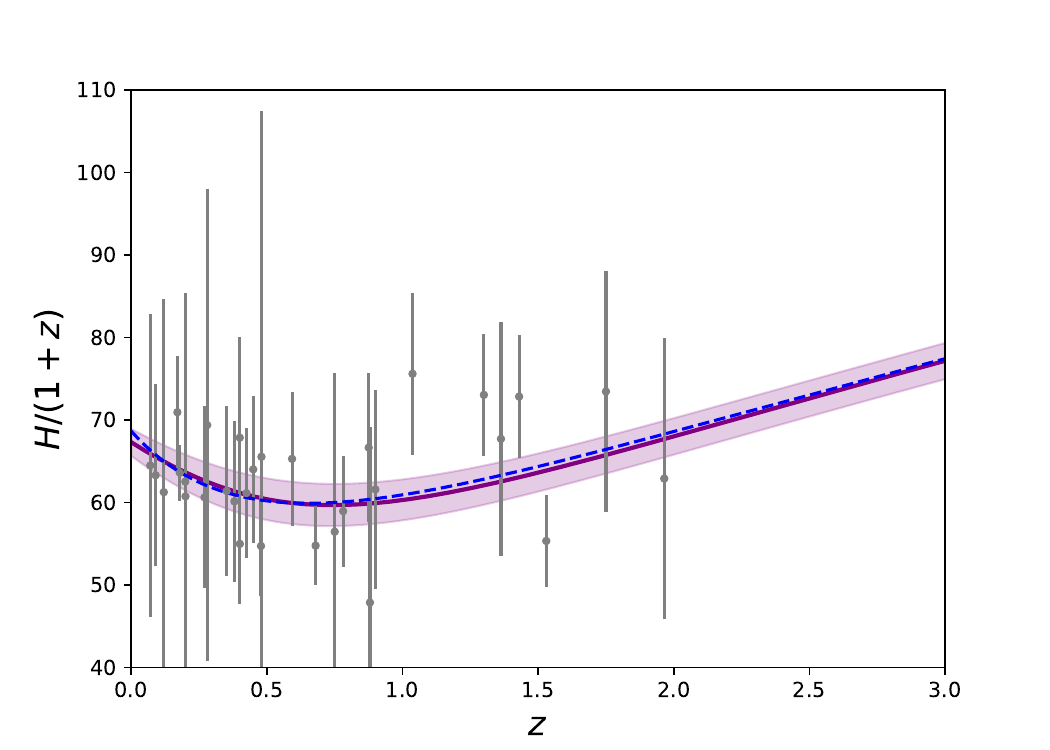}\includegraphics[scale=0.5]{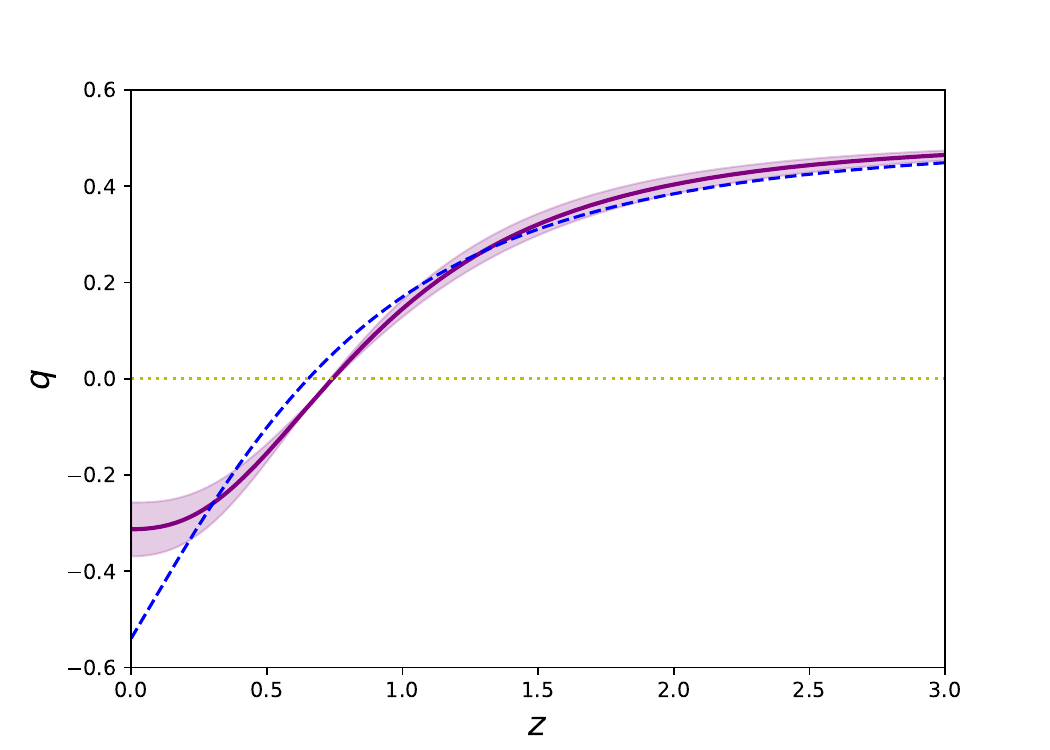}
	\caption{\label{fighubq} The behavior of the rescaled Hubble parameter $H(z)/(1+z)$ (left panel) and of the deceleration parameter $q(z)$ (right panel) as a function of the redshift $z$  for the Weyl Boundary gravity model for the best fit values of the parameters, as given by table \ref{bestfit}. The shaded area denotes the $1\sigma$ error. The dashed line represents the $\Lambda$CDM model.}
\end{figure*}

The behaviors of the Hubble and of the deceleration parameters are depicted in Fig.~\ref{fighubq}. From now on, we will use the the values of the parameters for the joint CC, Pantheon$^+$, BAO dataset. One can see from these Figures that the predictions of the Weyl Boundary cosmological model is almost the same as of the standard $\Lambda$CDM model. Slight differences from the $\Lambda$CDM model can be seen at redshifts $z\in(1,2)$,  where the Hubble parameter becomes smaller than that of the $\Lambda$CDM model. The deceleration parameter on the other hand predicts different value for the present time acceleration of the Universe. However, for redshifts bigger that $z\approx1$, the predictions of two models become more similar. In summary, the Weyl boundary gravity model  predicts a higher deceleration rate at earlier times and lower acceleration rate at late times, as compared to the $\Lambda$CDM model. 

\subsubsection{The cosmographic parameters}

The concordance between the theoretical predictions of the Weyl boundary cosmological model and the observations can also be further investigated by considering higher order derivatives of the Hubble parameter.  The cosmographic parameters are important (theoretically) observational quantities that can help in discriminating between various cosmological models. The Taylor series expansion of the scale factor can be generally represented as \cite{SJ}
\begin{align}
a(t)&=a_0\Big[1+H_0\left(t-t_0\right)-\frac{1}{2!}q_0H_0^2\left(t-t_0\right)^2\nonumber\\
&+\frac{1}{3!}j_0H_0^3\left(t-t_0\right)^3+\frac{1}{4!}s_0H_0^4\left(t-t_0\right)^4+\mathcal{O}(5)\Big],
\end{align}
where the jerk $j$ and snap $s$ parameters are defined as
\begin{align}
j=\frac{1}{H^3}\frac{1}{a}\frac{d^3a}{dt^3},\qquad s=\frac{1}{H^4}\frac{1}{a}\frac{d^4a}{dt^4}.
\end{align}

In terms of the deceleration parameter, $j$ and $s$ are obtained as \cite{SJ}
\begin{align}
j&=q+2q^2+(1+z)\frac{dq}{dz},\\
s&=-(1+z)\frac{dj}{dz}-2j-3jq,
\end{align}
respectively.

\begin{figure*}
	\includegraphics[scale=0.5]{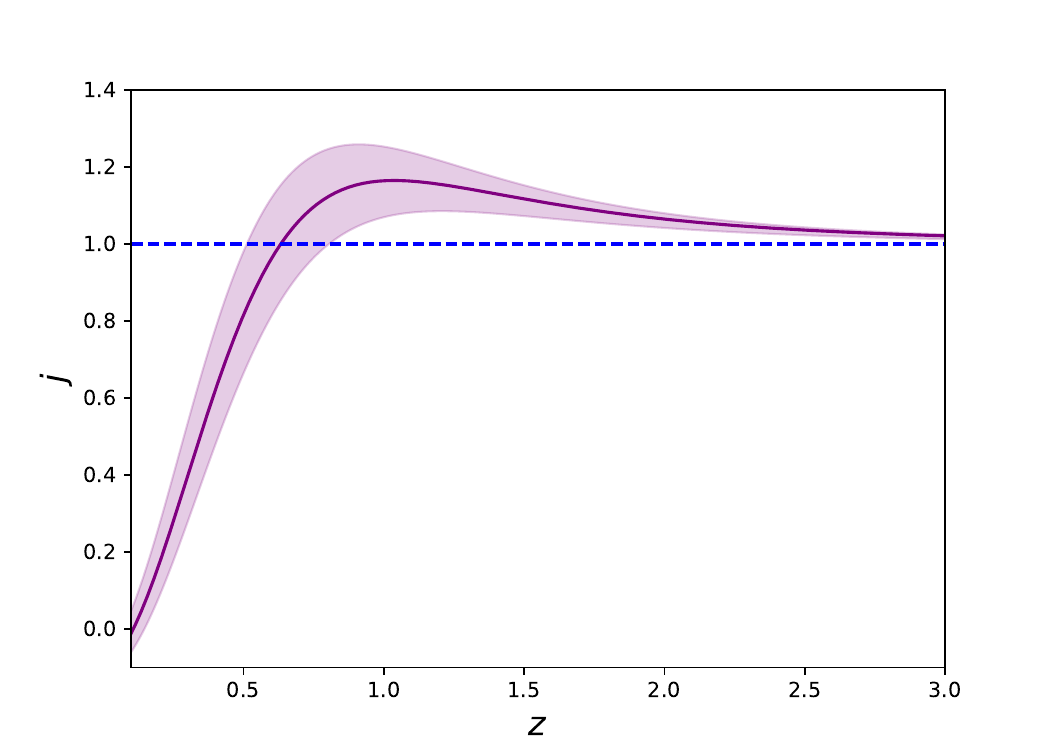}\includegraphics[scale=0.5]{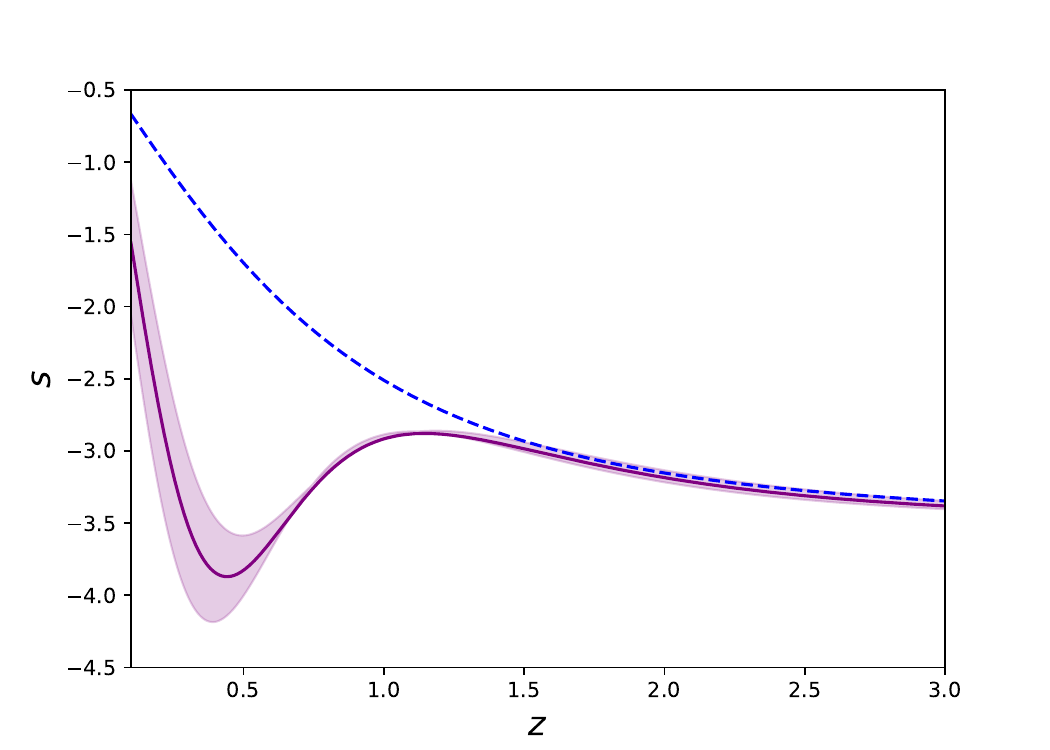}
	\caption{\label{figjerksnap} The behavior of the jerk parameter $j(z)$ (left panel) and of the snap parameter $s(z)$ (right panel) as a function of the redshift $z$  for the Weyl Boundary gravity model for the best fit values of the parameters, as given by table \ref{bestfit}. The shaded area denotes the $1\sigma$ error. The dashed line represents the $\Lambda$CDM model.}
\end{figure*}

\paragraph{The jerk and snap parameters.} The behaviors of the jerk and snap parameters are presented in Fig.~\ref{figjerksnap}. 
\begin{figure*}
	\includegraphics[scale=0.5]{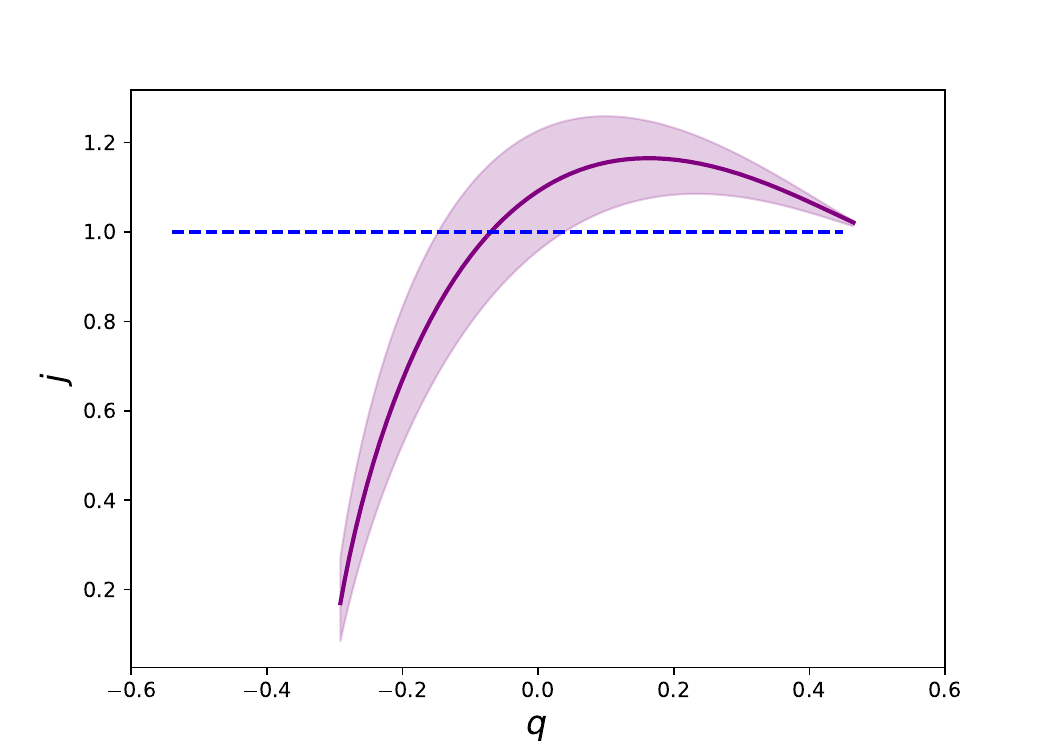}\includegraphics[scale=0.5]{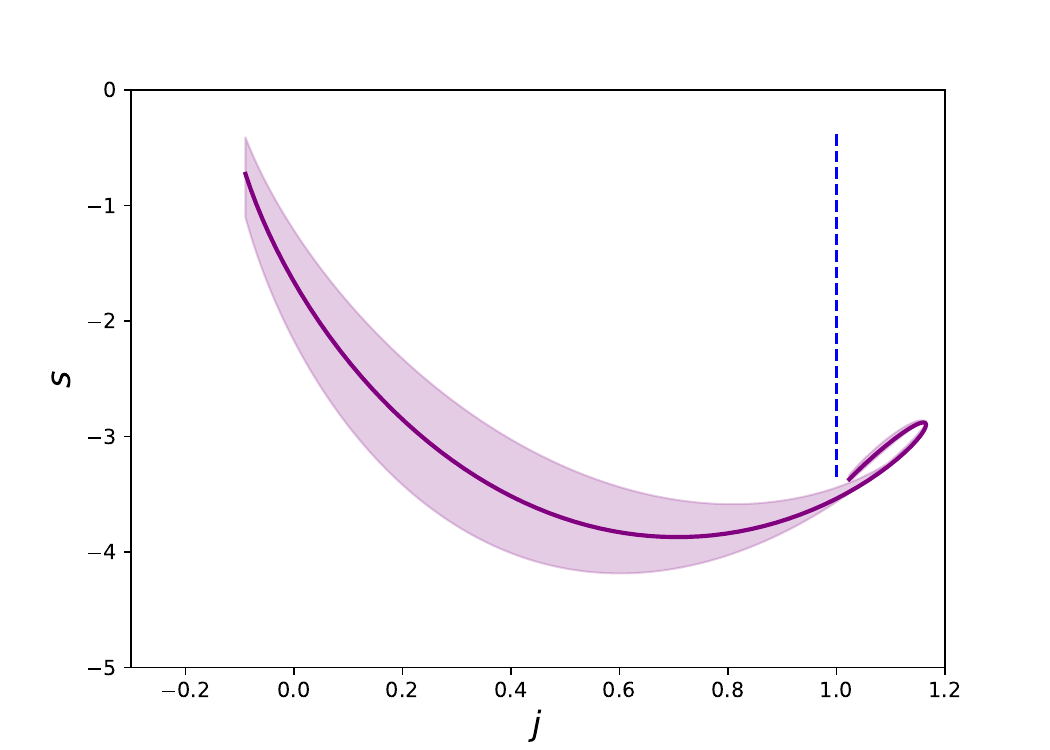}
	\caption{\label{figversus} The behavior of the jerk parameter $j$ as a function of the deceleration parameter $q$, $j=j(q)$ (left panel), and of the snap parameter as a function of the jerk parameter, $s=s(j)$ (right panel) for the Weyl Boundary gravity model for the best fit values of the parameters as given by table \ref{bestfit}. The dashed line represents $\Lambda$CDM model.}
\end{figure*}
\begin{figure*}
	\includegraphics[scale=0.5]{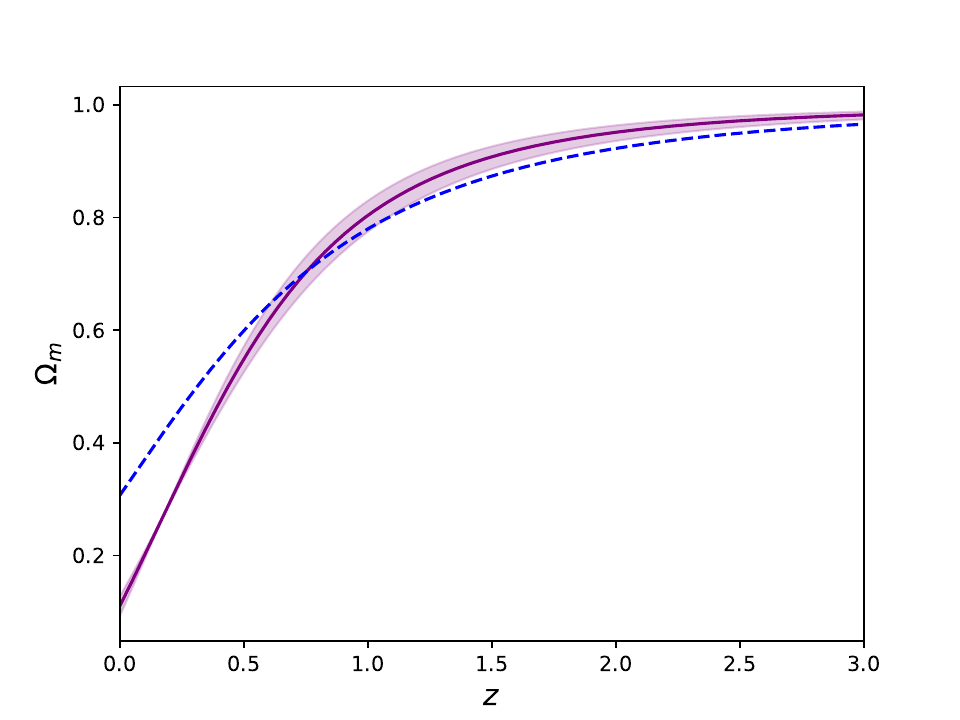}\includegraphics[scale=0.5]{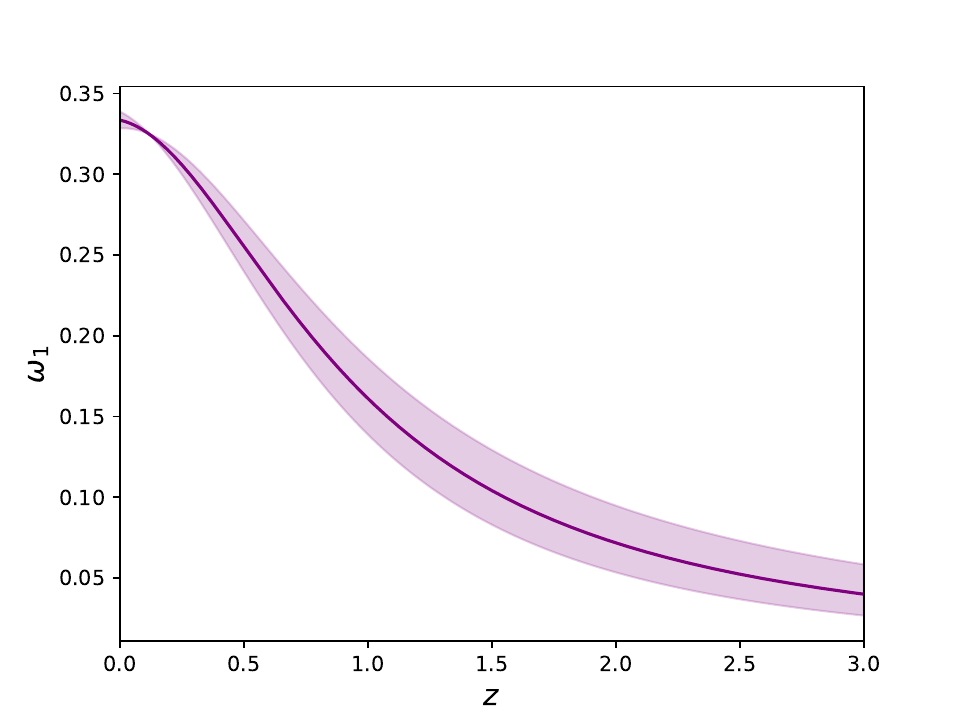}
	\caption{\label{OmegaandB} The behavior of the matter density abundance $\Omega_m$ and $\omega_1$ as functions of the redshift $z$ for the Weyl Boundary gravity model for the best fit values of the parameters as given by table \ref{bestfit}. The dashed line represents $\Lambda$CDM model.}
\end{figure*}
\begin{figure*}
	\includegraphics[scale=0.5]{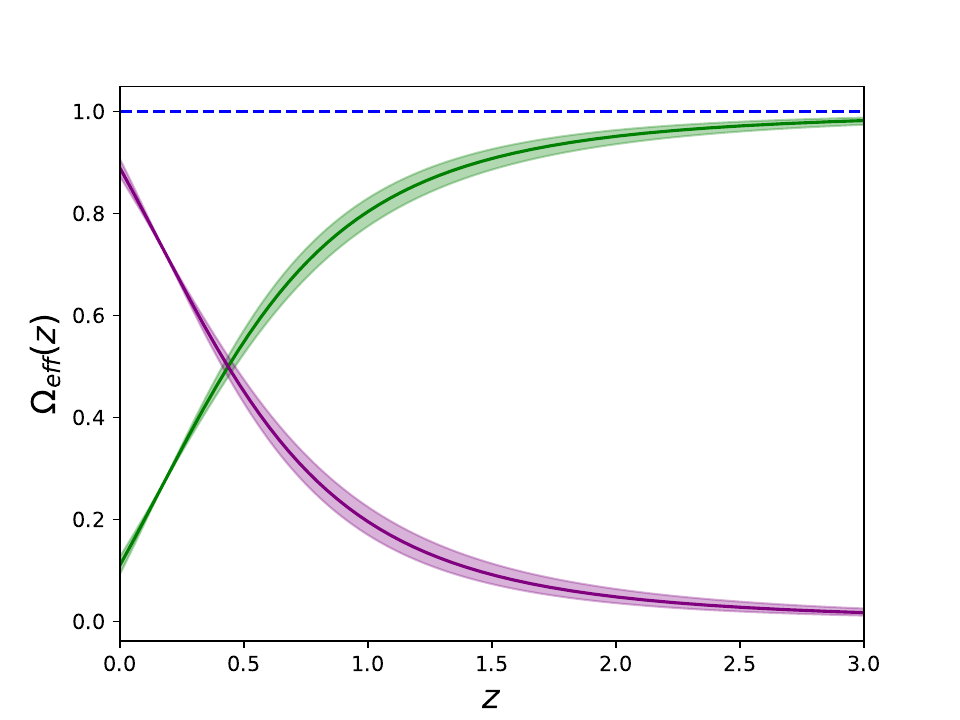}\includegraphics[scale=0.5]{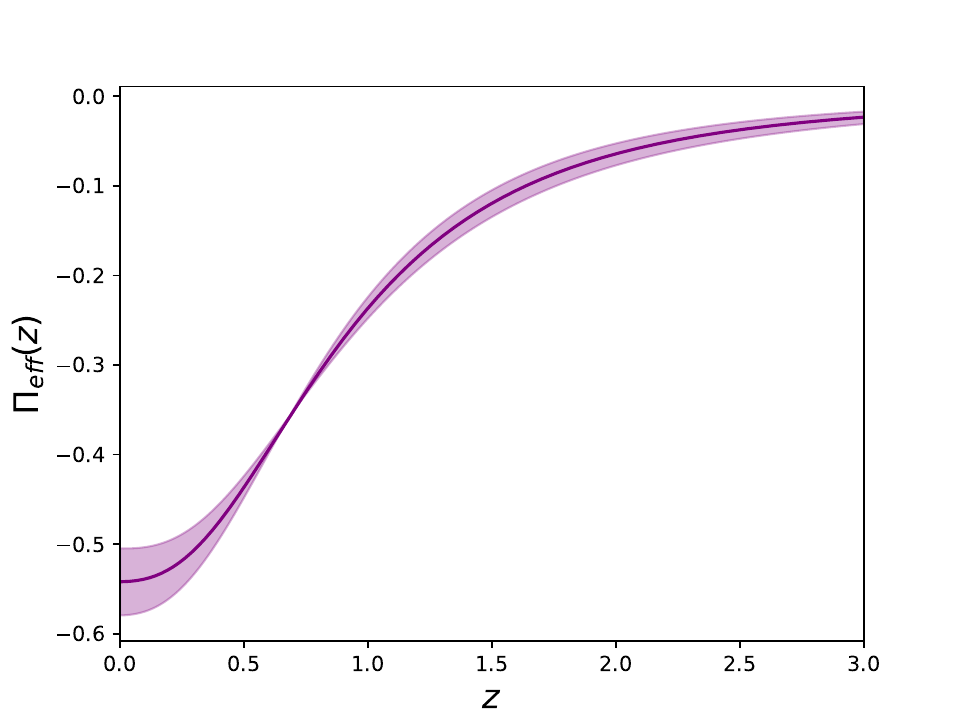}
	\caption{\label{figmattertotal} The behavior of the effective energy density abundance $\Omega_{eff}$ (left panel), and of the effective pressure abundance $\Pi_{tot}=\Pi_{eff}$ as a function of redshift $z$ (right panel) for the Weyl Boundary gravity model for the best fit values of the parameters as given by  table \ref{bestfit}. The dashed line represents the total matter abundance and the green curve represents the baryonic matter abundance also presented in figure \eqref{OmegaandB}.}
\end{figure*}
\begin{figure*}
	\includegraphics[scale=0.5]{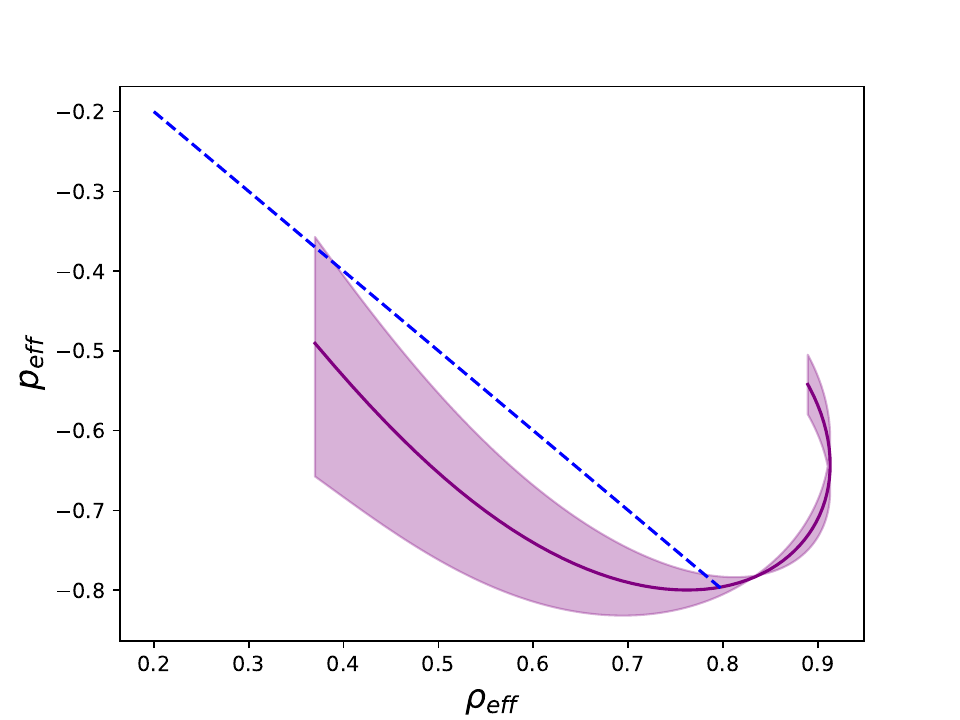}\includegraphics[scale=0.5]{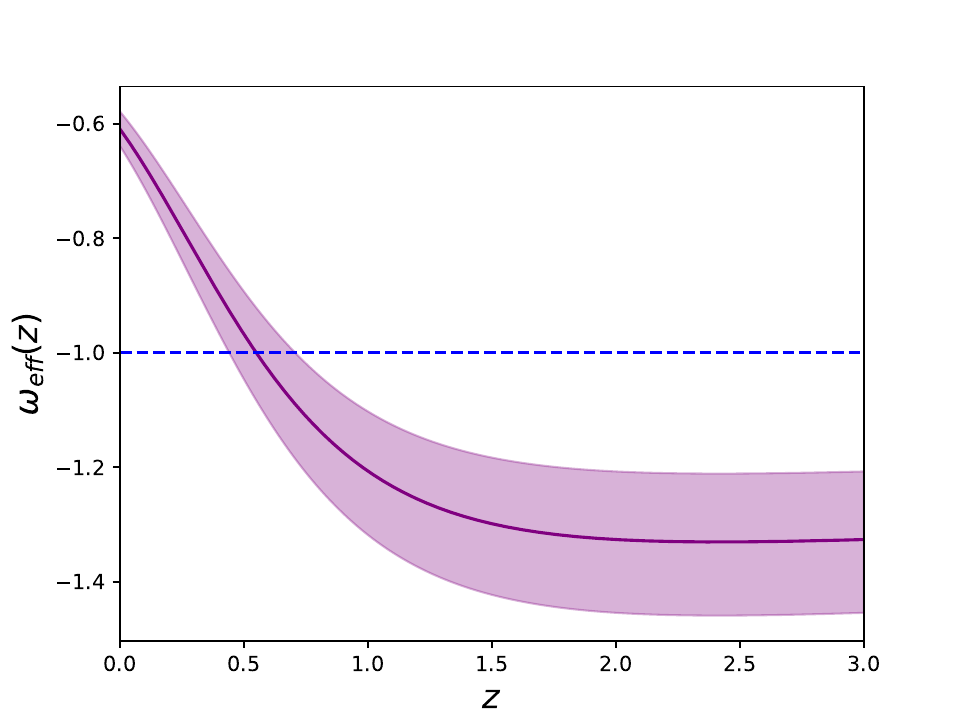}
	\caption{\label{figpvsrho} The behavior of the effective pressure $p_{eff}$ as a function of the effective energy density $\rho_{eff}$, and the equation of state parameter of the dark energy $\omega_{eff}=p_{eff}/\rho_{eff}$ as a function of redshift, for the Weyl Boundary gravity model for the best fit values of the parameters as given by Eqs.~(\ref{bestfit}). The shaded area denotes the $1\sigma$ error.  The dashed line represents $\Lambda$CDM model.}
\end{figure*}
From Fig.~\ref{figjerksnap} one can see that the jerk and snap parameters behave differently in the Weyl boundary cosmological model as compared to their $\Lambda$CDM counterparts. Noticing that the jerk parameter gives the convexity of the Hubble diagram, one can see that the convexity of the Weyl Boundary Hubble parameter increases with $z$ at late times, reaching to a maximum at around $z\approx0.7$, and then decreases to approach to the $\Lambda$CDM value. In the Hubble diagram one can also see this effect as the Weyl Boundary Hubble parameter is described by a more convex curve than its $\Lambda$CDM counterpart. In the behavior of the deceleration parameter one can see that the slope of $q$ is increasing at present times until $z\approx0.7$, where the increase rate starts to slow down, causing the deceleration curve to approach the $\Lambda$CDM behavior at earlier times.  The snap parameter, on the other hand, always lies below that $\Lambda$CDM value, gets negative values and reached the $\Lambda$CDM value at early times. This is also apparent from the behavior of the deceleration parameter, by noticing the negative convexity of the diagram.
\\ \indent
The variation of the jerk parameter in terms of the deceleration parameter, and of the snap parameter in terms of the jerk parameter are plotted in Fig.~\ref{figversus}. The differences between the Weyl Boundary model and the $\Lambda$CDM model can also be seen from these diagrams. 

\subsubsection{Matter and dark energy density parameters}

\paragraph{The matter density parameter.} In Fig.~\ref{OmegaandB} we have plotted the behavior the matter density parameters $\Omega_m = \bar\rho_m/h^2$ and $\omega_1$ as functions of the redshift $z$. As one can see from these Figures,  although the matter is not conserved in this model, the behavior is not very different from the $\Lambda$CDM case, represented by  the dashed line. 

As compared to the $\Lambda$CDM model, however, one can see that the Weyl Boundary model prefers lower amounts of matter in the energy budget of the Universe. However, the rate of change of the matter density parameter in the Weyl Boundary model is more than its $\Lambda$CDM counterpart, causing that at redshifts around $z\approx0.5$ the value of the matter density becomes larger than the $\Lambda$CDM value. 
At earlier times however, the value of the matter density parameter approaches the $\Lambda$CDM value. Also, the function $\omega_1$, which determines the Weyl Boundary effects in the model, is a decreasing function of the redshift, and approaches to zero at earlier times. 

The present-day value of the function $\omega_1$ can be obtained from Eq.~\eqref{omega1} and Table \ref{bestfit} as
\begin{align}
	\omega_1(z=0) = 0.333^{+0.0051}_{-0.0053}.
\end{align}

\paragraph{Dark energy density parameters.} In Fig.~\ref{figmattertotal} we have plotted the effective energy density and total pressure abundances, defined as
\begin{align}
	\Omega_{eff}=\frac{\kappa^2\rho_{eff}}{3H^2},\quad \Pi_{eff}=\frac{\kappa^2p_{eff}}{3H^2},
\end{align}
where we have used the definitions \eqref{rhoeff} and \eqref{peff}, and assumed in our calculations that $p=0$. 

One can see that the dark energy density parameter is a decreasing function of the redshift. This is expected, since in our definition we have $\Omega_{eff} = 1-\Omega_m$. The green curve in the Figure represents the matter density parameter, and the dashed line represents the total matter density parameter. The pressure abundance however is negative, and it is an increasing function of the redshift, approaching zero at earlier times. 

One can also see that the behavior of the energy density and pressure of the effective dark energy mimics the dark energy equation of state $p_{eff}\approx - \rho_{eff}$. 
In Fig.~\ref{figpvsrho}, we have plotted the behavior of the equation of state parameter of the dark energy, defined by
\begin{align}
	\omega_{eff} = \frac{p_{eff}}{\rho_{eff}},
\end{align}
as a function of redshift $z$. One can see that the equation of state parameter is a negative and decreasing function of the redshift, starting from values above $-1$, and tending to values around $-1.4$ at large $z$. This implies that the behavior of the dark energy fluid in this model is quintessence-like at late times and phantom-like at earlier times. Also, for redshifts $z\gtrsim1.5$, the equation of state parameter is almost constant, mimicking the cosmological constant at early times. In Fig.~\ref{figpvsrho}, we have also plotted the $p-\rho$ relation, which shows the behavior of the parameter of the equation of state in a clearer way.

\begin{figure}
	\includegraphics[scale=0.5]{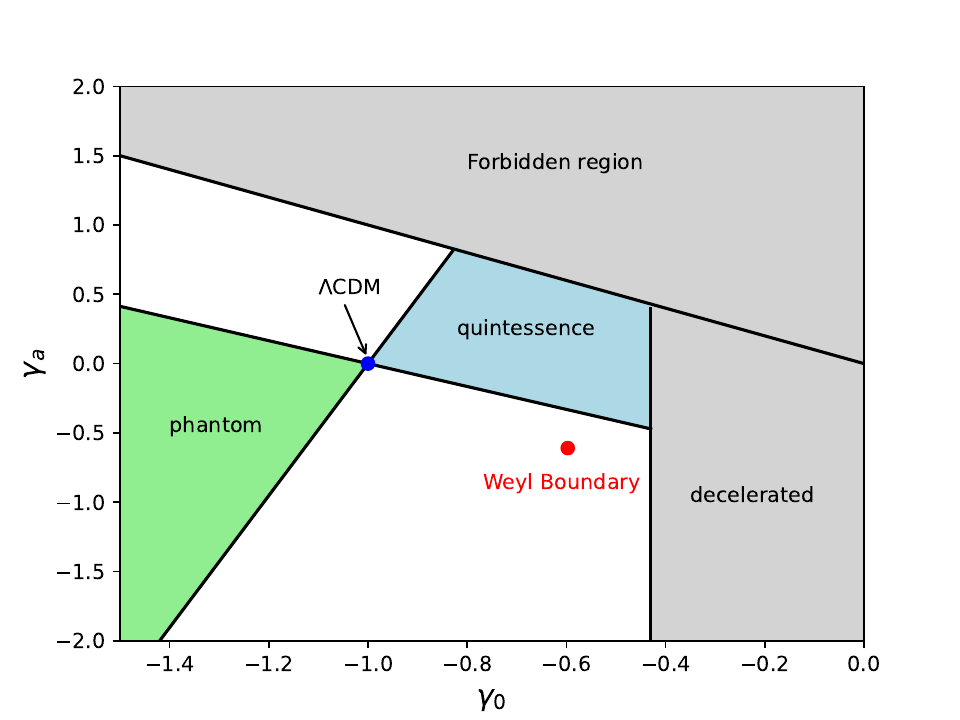}
	\caption{\label{baplot} Different regions described by the Barboza-Alcaniz parameterization. The Weyl Boundary model belongs to the white region which has a behavior that at some point the behavior of dark energy changes from quintessence to phantom.}
\end{figure}

As we have mentioned before, the behavior of the effective energy in the Weyl Boundary model is assumed to be similar to the Barboza-Alcaniz parametrization of dark energy. In Fig.~\ref{baplot}, we have plotted the different regions described by this parametrization, together with the behavior of the Weyl Boundary model. From the Figure, one can see that the Weyl Boundary model belongs to the white region, which indicates that at some point the dark energy changes from quintessence to phantom. This is in fact in line with the discussion we have made earlier about the effective energy density in Weyl Boundary model.

\begin{table}[h]
	\centering
	\begin{tabular}{|c|c||c|}
		\hline
		Model & $\ln \mathcal{Z}$ & $\ln B_{W\Lambda}$ \\ \hline\hline
		$\Lambda$CDM  & $-929.378 \pm 0.279$ & \multirow{2}{*}{$4.887 \pm 0.335$} \\ \cline{1-2}
		Weyl Boundary & $-924.490 \pm 0.279$ & \\ \hline
	\end{tabular}
	\caption{The evidences and Bayes factor for the $\Lambda$CDM and Weyl Boundary models.}
	\label{table3}
\end{table}
Let us now compare the Weyl Boundary model with the $\Lambda$CDM model by defining the Bayes factor as
\begin{align}
	B_{W\Lambda} = \frac{\mathcal{Z}_{WB}}{\mathcal{Z}_{\Lambda CDM}},
\end{align}
where $\mathcal{Z}_A$ is the marginal likelihood of model $A$. The Bayes factor quantifies how strongly the data favors one model over the other. Here we adopt the Jeffreys scale \cite{jeffrey} in which $|\ln B_{W\Lambda}|<1$ indicated inconclusive evidence, $1<|\ln B_{W\Lambda}|<2.5$ indicated weak evidence, $2.5<|\ln B_{W\Lambda}|<5$ corresponds to moderate evidence and $|\ln B_{W\Lambda}|>5$ indicates moderate evidence in favor of the model with higher evidence. With our definitions, positive result favors Weyl Boundary while negative result favors $\Lambda$CDM model. In Table~\ref{table3} we have summarized the result of this analysis, for both models. It is evident from these values that the Weyl Boundary model strongly favors over the $\Lambda$CDM model. This means that the data are fitted better in Weyl Boundary model and its predictions would be more reliable than those of $\Lambda$CDM model.

Let us now consider the tension between the value of the Hubble parameter inferred in this model compared to the value obtained from CMB observations. From the Planck 2018 results \cite{planck} one can read the value of the Hubble parameter as
$$ H_0 = 67.2733^{+0.6002}_{-0.5961}.$$
Now using the value of the Hubble parameter of model from table \ref{bestfit} one can obtain the tension between Planck 2018 data and the Weyl Boundary model which is summarized in Table \ref{H0comparison}.
\begin{table}
	\centering
	\begin{tabular}{|c||c|c|}
		\hline
		\textbf{Dataset} & \textbf{$H_0$} & \textbf{Tension} \\
		\hline\hline
		Planck 2018& $67.2733^{+0.60}_{-0.60}$ & -- \\
		\hline
		WB (CC+PNT+BAO) & $67.3085^{+1.6083}_{-1.6662}$ & $0.02\sigma$ \\
		\hline
		WB (CC+PNT)     & $67.3048^{+1.5655}_{-1.6030}$ & $0.02\sigma$ \\
		\hline
		WB (CC+BAO)     & $66.0899^{+1.6454}_{-1.7310}$ & $0.66\sigma$ \\
		\hline
		WB (CC)      & $65.2821^{+3.9669}_{-3.9435}$ & $0.50\sigma$ \\
		\hline
		$\Lambda$CDM (CC+PNT+BAO) & $68.6525^{+1.6251}_{-1.6307}$ & $0.80\sigma$ \\
		\hline
	\end{tabular}
	\caption{\label{H0comparison} The tension between Planck 2018 data and the Weyl Boundary model using four difference dataset combinations. We have also included the result for $\Lambda$CDM model.}
\end{table}
It is evident that the tension of Weyl Boundary model with the CMB data is about $0.28\sigma$ which indicates that the present model is in full agreement with the Planck observations. It should be noted here that the well-known $H_0$ tension is between the predictions of Planck data and late-time observations with SH0ES calibration \cite{SH0ES}. Here, we have ignored this tension by inferring the value of absolute magnitude $\mathcal{M}$ directly from the cosmological observations, instead of using the SH0ES calibration. As a result, there is no Hubble tension present here. As a summary our dataset choice does not test the SH0ES–Planck discrepancy because it does not incorporate the local distance-ladder calibration.

\subsection{$Om(z)$ diagnostic} 
The $Om(z)$ diagnostic tool is an important theoretical method allowing to estimate the deviations of the alternative cosmological models from the standard  $\Lambda$CDM paradigm. The $Om(z)$ function can be used to determine the physical nature of a given dark energy model, and to infer if the cosmological fluid  is a phantom-like, quintessence-like, or it can be described by a cosmological constant.

The main quantity of the $Om(z)$ diagnostic is the $Om(z)$ function, defined as \cite{Om}
\begin{equation}\label{om}
	Om(z)=\frac{h^{2}(z)-1}{(1+z)^{3}-1}.
\end{equation}

For the standard $\Lambda$CDM model, the function $Om(z)$ has a very simple form, being a universal constant, equal to the present day matter density parameter $\Omega_{m0}$. For cosmological models with a constant parameter of the equation of state of dark energy, $w={\rm constant}$, a phantom-like behavior is indicated by a positive slope of $Om(z)$. On the other hand, a negative slope corresponds to  a quintessence-like evolution.  For  $w$CDM cosmology, by assuming that the dynamical dark energy fluid can be described by  a linear barotropic equation of state, with the parameter of EOS denoted by $w$, one can find
\be
Om(z)=\frac{\Omega_{m 0}(1+z)^{3}+\left(1-\Omega_{m 0}\right)(1+z)^{3(1+w)}-1}{(1+z)^{3}-1} .
\ee

\begin{figure}[tbp]
	\centering
	\includegraphics[scale=0.55]{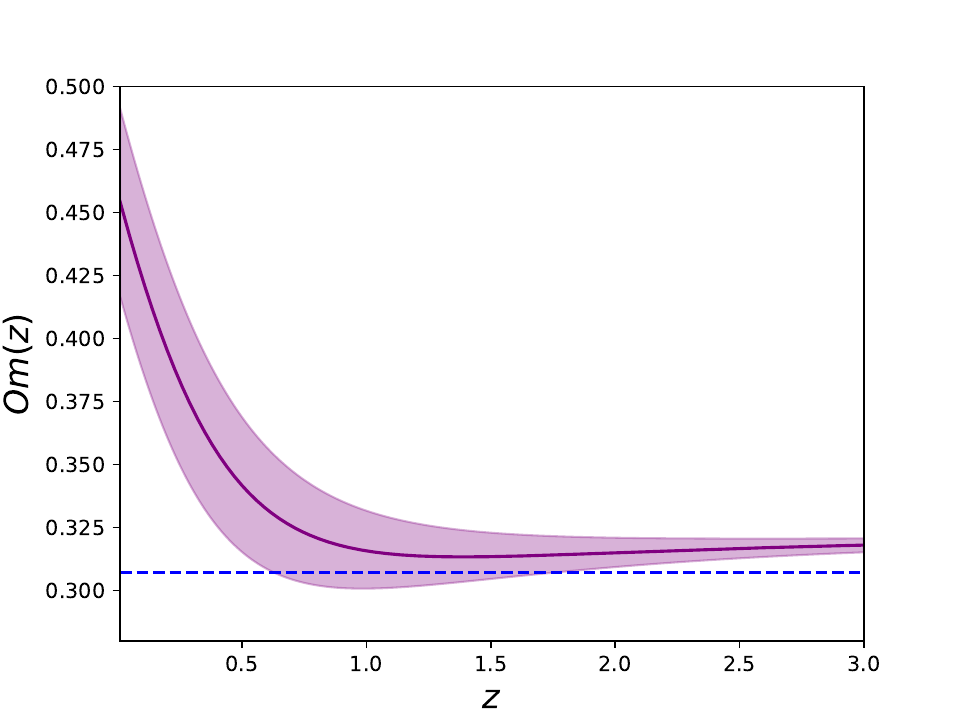}
	\caption{\label{Omdiag} The redshift variation of the $Om(z)$ diagnostic function for the Weyl Boundary geometric gravity model for the best fit values of the parameters as given by Eqs.~(\ref{bestfit}). The shaded area denotes the $1\sigma$ error. The dashed line represents the $\Lambda$CDM model.}
\end{figure}

In the case of the $\Lambda \mathrm{CDM}$ model, the parameter of the dark energy equation of state is $w=-1$, and we obtain
$Om(z)=\Omega_{m 0}$.
The variation with respect to the redshift of the $Om(z)$ diagnostic function for the Weyl Boundary geometric gravity cosmological model is represented in Fig.~\ref{Omdiag}. Noticing the negative slope of the $Om$ diagram in the range $z\in(0,1.5)$, one can deduce that the effective fluid in the Weyl Boundary model is quintessence-like. However, as can be seen from the Figure, for larger values of the redshift $z\gtrsim1.5$ the $Om$ diagram has a positive slope indicating that the dark energy behaves like phantom. 

It should be noted however that the value of the $Om(z)$ function is very close to the current value of the matter density parameter, which means that the theory mimics $\Lambda$CDM model at earlier times. These results are compatible with our discussions in the previous Sections. 

One should also note that the equation of state parameter of dark energy in the Weyl Boundary model is not exactly constant and as a result the qualitative description of the $Om(z)$ diagnostic is not exactly the same as $\omega$CDM model discussed above. As a result, the exact value of the redshift is not reliable but the quintessence-phantom nature of dark energy can be inferred from the analysis, in line with our previous discussions on the model.

 \section{Conclusions and final remarks}\label{concl}
 
 In the present work we have investigated the cosmological implications of a generalization of the Einstein-Hilbert action of standard general relativity, by including into the general mathematical structure of the theory the contribution of the boundary terms, which are generally neglected in the study of the gravitational dynamics. With respect to the properties of the boundary, we have assumed that its properties are of Weyl geometric type. A general mathematical results is that the variation of the connection with respect to the metric 
is obtained with the help of the covariant derivative $\nabla _\lambda$ of the variation $\delta g^{\mu \nu}$ of the metric tensor. 

In order to implement the idea of a Weylian boundary we have substituted  the Riemannian covariant derivative with its Weylian counterpart, so that $\nabla _\lambda \rightarrow \tilde{\nabla}_\lambda$, where $\tilde{\nabla}_\lambda$ is the Weylian covariant derivative. This substitution allows us to use extensively the formalism of Weyl geometry in the description of the boundary terms, which leads to obtaining the contribution of the Weylian boundary to the Einstein-Hilbert action as given by Eq.~(\ref{40}). Thus, the Weylian boundary term is a function of the Weyl vector, and of its covariant derivative, and it gives a non-zero contribution to the field equations. 

As an application of the gravitational model introduced in this work we have explored its cosmological implications for the evolution of the late Universe. We have obtained the generalized Friedmann equations (\ref{Fr1}) and (\ref{Fr2}), which contain the contribution of the Weylian boundary in the form of an effective energy density and pressure, of geometric origin, and which can be interpreted as a geometric type dark energy. However, since the theory is not derived directly from an action, the matter conservation does not hold in this model. Also, the Weyl vector field comes from the boundary and as a result there is no equation of motion for it. In order to fix the dynamics of the Weyl vector field, we have assumed the Barboza-Alcaniz parametrization for the Weyl Boundary effective dark energy type terms.

To test the validity of the Weylian type boundary cosmology we have compared its predictions with the joint observational data of the Hubble parameter from the Cosmic Chronometer method, together with the Pantheon$^+$ dataset without SH0ES calibration and the BAO measurements from the DESI DR2 dataset, by performing an MCMC type statistical analysis, which led to the determination of the model parameters. Overall, we have found a very good concordance between the predictions of the model, the observational data, and the $\Lambda$CDM model at the level of the Hubble and deceleration parameters up to a redshift of $z\approx 3$. The $Om(z)$ diagnostic function also shows that the Weylian boundary model has a quintessence to phantom transition.  We have also seen that for large values of $z$ the Weyl vector tends to zero. Hence, the effects of the Weylian boundary are negligibly small at least for a certain period of evolution in the early Universe, but they become more and more important during the late cosmological time evolution. 

An interesting question is the existence of a de Sitter type solution in this model. For a vacuum state with $\rho=p=0$, and $H=H_0={\rm constant}$, the Friedmann equations (\ref{Fr1}) and (\ref{Fr2}), together with the conservation equation give for the Weyl vector the set of equations
\begin{align}
\dot{\omega}_0&=\omega _0\left(H_0+\alpha \omega _0\right),\nonumber\\
\dot{\omega}_0&=-H_0\left(H_0+\alpha \omega _0\right).
\end{align}
Combining these equations, one obtains         
\be
\omega _0(t)=-H_0.
\ee
Hence, the model admits a de Sitter type solution in the conservative case only for a constant Weyl vector. 

 In the present work we have proposed a geometric approach to the description of the gravitational dynamics, which, on one hand, generalizes the Einstein-Hilbert action, and, on the other hand, combines in a single framework of Riemann and Weyl geometries, which may coexist on cosmic scales.  The  obtained results may lead to the development of some basic mathematical and theoretical methods that could help in the understanding of the dynamical laws and processes that govern the evolution of the  Universe.
 
 \section{Acknowledgments}
 
 We would like to thank the anonymous referee for comments and suggestions that helped us to improve our manuscript.


\begin{thebibliography}{99}

\bibitem{GH} G. W. Gibbons and S. W. Hawking,  Phys. Rev. D \textbf{15}, 2752 (1977).

\bibitem{Y} J. W. York, Foundations of Physics \textbf{16}, pages 249 (1986).

\bibitem{Bo1} C. G. Boehmer and E. Jensko, Phys. Rev. D {\bf 104}, 024010 (2021).

\bibitem{Bo2} C. G. Boehmer and E. Jensko, J. Math. Phys. {\bf 64}, 082505 (2023).

\bibitem{Bo3} C. G. Boehmer and A. d'Alfonso del Sordo, General Relativity and Gravitation {\bf 56}, 75 (2024). 

\bibitem{BR1} J. S. Ridao, M. Bellini, Astrophys. Space Sci. {\bf  357}, 94 (2015).

\bibitem{BR2} J. S. Ridao, M. Bellini, Phys. Lett. B \textbf{751},  565 (2015).

\bibitem{B3} R. Hernandez-Jimenez, C. Moreno, M. Bellini, C. Ortiz, Phys. Dark Univ. \textbf{38},  101137 (2022).

\bibitem{fq1} J. M. Nester and H.-J. Yo, Chin. J. Phys. {\bf 37}, 113 (1999).

\bibitem{fq2}  J. Beltran Jimenez, L. Heisenberg, and T. S. Koivisto, JCAP {\bf 08}, 039 (2018).

\bibitem{fq3} J. Beltran Jimenez, L. Heisenberg, and T. Koivisto, Phys. Rev. D 98, 044048 (2018).

\bibitem{fq4}  L. Heisenberg, Phys. Rept. {\bf 1066}, 1 (2024).

\bibitem{DLS} A. De, T.-H. Loo, and E. N. Saridakis, Journal of Cosmology and Astroparticle Physics {\bf  2024},  03, 050 (2024).

\bibitem{fQC1} S. D. Sadatian and S. M. R. Hosseini, Physics of the Dark Universe {\bf 47}, 101737 (2025). 

\bibitem{fQC2} A. Paliathanasis, New Astronomy {\bf 120}, 102426 (2025).

\bibitem{fQC3} G. Murtaza, A. De, T.-H. Loo,  Y. K. Goh, and H. H. Liew, Annals of Physics {\bf 480}, 170086 (2025).

\bibitem{fQC4} S. R. Bhoyar and Y. B. Ingole, New Astronomy {\bf 118}, 102386 (2025).

\bibitem{fQC5} A. Samaddar, and S. S. Surendra,  Journal of High Energy Astrophysics {\bf 48}, 100404 (2025).

\bibitem{fQC6} Shaily, J. K. Singh, M. Tyagi, and J. R. L. Santos, Physics of the Dark Universe {\bf 48}, 101946 (2025).

\bibitem{fQC7} A. Samaddar and S. S. Singh, Physics of the Dark Universe {\bf 47}, 101792 (2025). 

\bibitem{CFF} S. Capozzielloa, V. De Falco and C. Ferrara, 	Eur. Phys. J. C {\bf 83}, 915 (2023).

\bibitem{Kad} S. A. Kadam, N. P. Thakkar, and B. Mishra, Eur. Phys. J. C  {\bf 83}, 809 (2023).

\bibitem{Pal} A. Paliathanasis, Physics of the Dark Universe {\bf 43},  101388 (2024).

\bibitem{Loh} S. V. Lohakare and B. Mishra, The Astrophysical Journal {\bf 978}, 26 (2025).

\bibitem{Harko} T. M. Matei and T. Harko, Physics of the Dark Universe {\bf 46}, 101578 (2024). 

\bibitem{LaLi} L. D. Landau and E. M. Lifshitz, The Classical Theory of
Fields, Butterworth-Heinemann, Oxford, 1999

\bibitem{Weyl}  H. Weyl, Sitzungsberichte der K\"{o}niglich Preussischen Akademie der Wissenschaften zu Berlin, 465 (1918).

\bibitem{Ghil} D. M. Ghilencea, Eur. Phys. J. C {\bf 83}, 176 (2023).

\bibitem{Ghil1} D. M. Ghilencea, The European Physical Journal C {\bf  83}, 176 (2023).

\bibitem{Ghil2} C. Condeescu, D. M. Ghilencea, and A. Micu, European Physical Journal C {\bf 84}, 292 (2024).

\bibitem{Ghil3} D. M. Ghilencea, Physical Review D {\bf 111}, 085019 (2025). 

\bibitem{Del} A. Delhom, I. P. Lobo, G. J. Olmo, and C. Romero, Eur. Phys. J. C {\bf 79},  878, (2019). 

\bibitem{WIG} C. Romero, J.B. Fonseca-Neto, and M. L. Pucheu, Class. Quantum Grav. {\bf 29}, 155015 (2012)
\bibitem{barbozaalcaniz} E. M. Barboza Jr. and J. S. Alcaniz, Phys. Lett. B 666, 415 (2008).
\bibitem{CCdata} A. Favale, A. Gómez-Valent and M. Migliaccio, MNRAS  523, 3406 (2023).
\bibitem{PANdata} D. Brout \textit{et al.}, ApJ,  938, 110 (2022).
\bibitem{noshoes} W. L. Freedman, B. F. Madore, I. S. Jang, T. J.
Hoyt, A. J. Lee, and K. A. Owens, arXiv:2408.06153; L. Perivolaropoulos, Phys. Rev. D 110, 123518 (2024).
\bibitem{bao} DESI Collaboration: M. Abdul-Karim, et. al, arXiv:2503.14738 [astro-ph.CO].
\bibitem{SJ} F. Y. Wang, Z. G. Dai, and S. Qi, A\& A {\bf 507}, 53 (2009).
\bibitem{Om} V. Sahni, A. Shafieloo, and A. A. Starobinsky, 2008, Phys. Rev. D {\bf 78}, 103502 (2008).
\bibitem{jeffrey}  H. Jeffreys, ``The theory of probability'', Oxford University Press 1961.
\bibitem{planck}
N. Aghanim et al. (Planck Collaboration), Astron. Astrophys. 641 (2020) A6.
	\bibitem{SH0ES}
A. G. Riess et al. (SH0ES Collaboration), Astrophys. J. Lett. 934 (2022) L7.
\end{thebibliography}
\end{document}